\documentclass[acmlarge]{acmart}

\usepackage{wrapfig}
\usepackage{gensymb}
\usepackage[linesnumbered,ruled,vlined]{algorithm2e}
\AtBeginDocument{%
  }

\setcopyright{acmcopyright}
\copyrightyear{2024}
\acmYear{2024}
\acmDOI{XXXXXXX.XXXXXXX}

\acmJournal{IMWUT}
\acmVolume{0}
\acmNumber{0}
\acmArticle{0}
\acmMonth{0}

\begin{document}

\title{UbiPhysio: Support Daily Functioning, Fitness, and Rehabilitation with Action Understanding and Feedback in Natural Language}



\author{Chongyang Wang}
\orcid{0000-0002-9819-088X}
\email{wangchongyang@tsinghua.edu.cn}
\affiliation{%
  \institution{Tsinghua University}
  \country{China}
}

\author{Yuan Feng}
\orcid{0000-0001-8237-0037}
\affiliation{%
  \institution{MPT, Department of Rehabilitation Medicine, West China Hospital, Sichuan University}
  \country{China}
}

\author{Lingxiao Zhong}
\orcid{0009-0002-1352-1879}
\affiliation{%
  \institution{Tsinghua University}
  \country{China}
}

\author{Siyi Zhu}
\orcid{0000-0001-8213-7622}
\authornotemark[2]
\email{hxkfzsy@scu.edu.cn}
\affiliation{%
  \institution{Department of Rehabilitation Medicine, West China Hospital, Sichuan University}
  \country{China}
}

\author{Chi Zhang}
\orcid{0009-0004-0846-4755}
\author{Siqi Zheng}
\orcid{0009-0002-6787-4223}
\author{Chen Liang}
\orcid{0000-0003-0579-2716}
\author{Yuntao Wang}
\orcid{0000-0002-4249-8893}
\email{yuntaowang@tsinghua.edu.cn}
\affiliation{%
  \institution{Tsinghua University}
  \country{China}
}

\author{Chengqi He}
\orcid{0000-0002-5349-0571}
\authornotemark[2]
\email{hxkfhcq2015@126.com}
\affiliation{%
  \institution{Department of Rehabilitation Medicine, West China Hospital, Sichuan University}
  \country{China}
}

\author{Chun Yu}
\orcid{0000-0003-2591-7993}
\authornote{Computer Science Corresponding Author. $^\dagger$Clinical Corresponding Author.}
\email{chunyu@tsinghua.edu.cn}
\affiliation{%
  \institution{Tsinghua University}
  \country{China}
}

\author{Yuanchun Shi}
\orcid{0000-0003-2273-6927}
\email{shiyc@tsinghua.edu.cn}
\affiliation{%
  \institution{Tsinghua University and Qinghai University}
  \country{China}
}

\renewcommand{\shortauthors}{C. Wang et al.}

\begin{abstract}

    We introduce UbiPhysio, a milestone framework that delivers fine-grained action description and feedback in natural language to support people's daily functioning, fitness, and rehabilitation activities. This expert-like capability assists users in properly executing actions and maintaining engagement in remote fitness and rehabilitation programs. Specifically, the proposed UbiPhysio framework comprises a fine-grained action descriptor and a knowledge retrieval-enhanced feedback module. The action descriptor translates action data, represented by a set of biomechanical movement features we designed based on clinical priors, into textual descriptions of action types and potential movement patterns. Building on physiotherapeutic domain knowledge, the feedback module provides clear and engaging expert feedback. We evaluated UbiPhysio's performance through extensive experiments with data from 104 diverse participants, collected in a home-like setting during 25 types of everyday activities and exercises. We assessed the quality of the language output under different tuning strategies using standard benchmarks. We conducted a user study to gather insights from clinical physiotherapists and potential users about our framework. Our initial tests show promise for deploying UbiPhysio in real-life settings without specialized devices.
    
\end{abstract}



\received{15 August 2023}
\received[revised]{15 November 2023}
\received[accepted]{12 January 2024}


\maketitle

\begin{figure}[ht]
  \centering
  \includegraphics[width=0.85\linewidth]{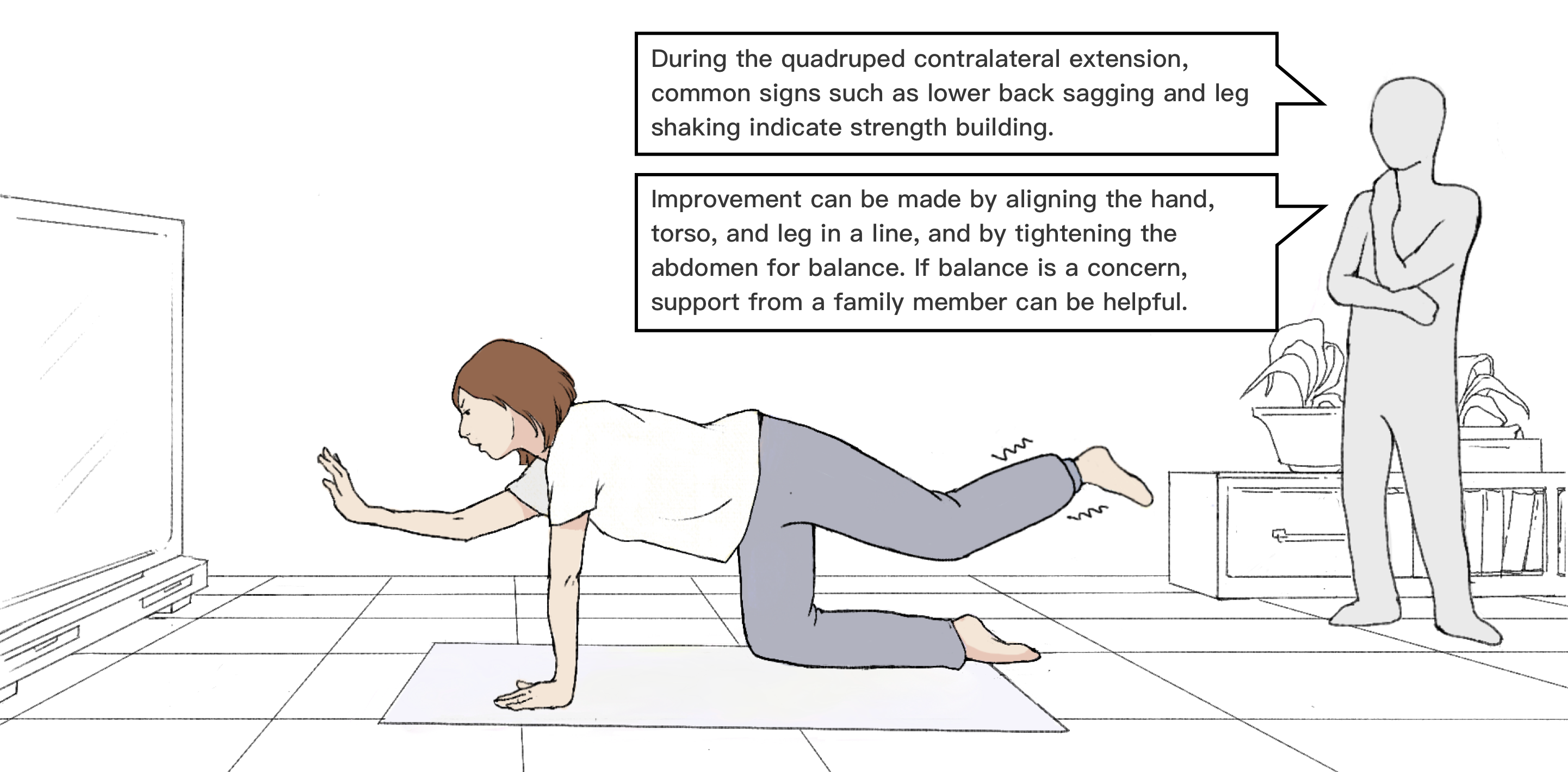}
  \caption{By accurately understanding the detailed movement patterns of a person, a physio is able to provide guidance on how to make the exercise more effective and suggestions that could help them achieve the goal properly. Inspired by such vivid and expert interactions, this work aims to narrow the gap between ubiquitous technology and the basic skill of a physio in evaluating one's action and generating detailed feedback.}
  \Description{}
  \label{fig:fig1}
\end{figure}

\section{Introduction}

    People's domestic activities and exercise routines provide vital insights for health providers to understand their difficulties, problems, and behavioral changes \cite{cook2009ambient}. Recently, home-based physiotherapy concerning everyday functioning and regular exercise has gained increasing importance, given the benefits they provide for enhancing individuals' quality of life \cite{piwek2016rise}, maintaining functional independence \cite{garber2011quantity}, combating chronic diseases \cite{law2014effects}, and promoting recovery from illness or injury outside hospital settings \cite{major2021feasibility}. However, without guidance, individuals can easily develop improper habitual behaviors that pose long-term musculoskeletal risks. For instance, bending to pick up a heavy bag could significantly harm the lower back, whereas squatting and then lifting the bag is a safer approach. Moreover, performing physical exercises without professional supervision can result in ineffective training or even injury due to incorrect postures and movements. Another concern is the lack of timely feedback in domestic settings, which can discourage those with chronic diseases from adhering to self-organized rehabilitation programs, especially given the high cost of clinical services \cite{mclean2010interventions,gutierrez2022effectiveness}. Recent advances in action captioning \cite{TM2T,T2M-GPT,jiang2023motiongpt,zhang2023motiongpt,haresamudram2023towards} and knowledge-driven natural language generation (NLG) \cite{guu2020retrieval,chen2022harnessing,du2022glm,ram2023context} present an opportunity to create a system that can perceive, understand, and offer expert-like feedback on home-based daily functional activities and exercises. Compared to existing works, our approach delves into fine-grained action descriptions that encompass both the type of action and potential movement patterns that could inform intervention. Furthermore, we introduce a simple yet efficient retrieval-based method that leverages the common-sense knowledge of a large language model (LLM) to generate professional feedback. 

    As illustrated in Fig~\ref{fig:fig1}, we receive motivations from the real life scenario, where a physio provides direct feedback on conducting an action properly and making the exercise doable and effective, grounded in a fine-grained understanding of the user's performance. In light of this, we ask: \textbf{how can technology support people's daily functioning and exercises with expert-like action understanding and feedback}? We introduce UbiPhysio, the first framework of its kind designed to function as a virtual physiotherapist, facilitating daily functional activities and supporting rehabilitation exercises. UbiPhysio accepts action data as input and connects it to natural language through self-supervised tokenization. In contrast to previous approaches, we propose employing biomechanical movement features and action-conditioned instruction tuning to deliver detailed action descriptions. Ultimately, leveraging the common-sense knowledge of a LLM, the framework generates vivid and professional feedback using a retrieval-enhanced prompting strategy.
    
    Our experiment leverages data collected from 104 participants, capturing full-body action data while performing 25 different types of daily functioning and exercise activity. We established the experiment with an aim to reveal the performance of participants at their homes. We achieved this by converting the lab space into a furnished, home-like environment, and implementing a continuous, less intrusive experimental session. To evaluate the efficacy of UbiPhysio, we carried out rigorous cross-validation tests on unseen participants, measuring the accuracy of the system's language output with standard objective benchmarking tools. Additionally, we conducted subjective evaluations with our participants and clinical physios. The results strongly suggest that UbiPhysio has the potential to be successfully integrated into real-life scenarios, paving the way for future advancements in ubiquitous fitness and rehabilitation technology. The main contributions of this work are summarized below. 

    \begin{itemize}
        \item We introduce UbiPhysio, a milestone framework capable of delivering expert-like action descriptions in terms of the action type and movement patterns-of-interest in natural language. It further generates personalized and professional feedback to improve user engagement and performance.
        \item To enhance the framework, we propose two novel tuning strategies, including the integration of comprehensive biomechanical bodily features for precise action quantification and the instruction tuning conditioned by the action type to optimize language output results. 
        \item We evaluate the framework with data collected from 43 healthy participants (22 males, 21 females, with an average age of 28.125 (std: 8.951)) and 61 participants with chronic lower-back pain (20 males, 41 females, with an average age of 32.582 (std: 10.433)) in a furnished, home-like environment. In addition to using benchmark metrics, we conduct a user study with the participants and physios to gain realistic insights on language outputs. Aside from using IMUs for full-body pose estimation by default, we test the framework using pose inputs acquired with a visual system, to demonstrate its potential for real-life deployment.
    \end{itemize}

\section{Related Work}

    This section provides a literature review on relevant studies on action-centric applications, as well as multimodal language modeling with action data.

\subsection{Analyzing Actions for Healthcare, Assistance, and Fitness}
    
    Recognizing the type or category of an action serves as the foundation of many ubiquitous systems and applications. However, human perception of actions extends beyond mere categorization, and are capable of inferring the emotional and physical status, needs, and difficulties, based on detailed action patterns \cite{giese2003neural,barrett2011context,aviezer2012body}. Therein, the difference between a layperson and a domain expert lies in their ability to \textbf{accurately capture such patterns and respond with proper language}. This capability is crucial for applications where technology is expected to play the role of a specialist in relevant domains \cite{xie2021hearfit+}.
    
    This work differs significantly from many previous studies focusing on activities of daily living (ADLs) \cite{hiremath2022bootstrapping,bhattacharya2022leveraging,moschetti2017daily}, assisted fitness \cite{hoang2016onebody,guo2018device,jo2023flowar,clarke2020reactive}, and remote rehabilitation \cite{su2014kinect,halloran2019remote}. Most of these studies aim to identify discrete activity classes and provide general insights, such as revealing behavioral changes (e.g., a decrease in exercise) through longitudinal observations, detecting critical or abnormal activities (e.g., falls or prolonged bed rest) \cite{cook2012casas}, and assessing health risks by tracking specific activities like drinking and eating \cite{burke2011self, plotz2011activity}. Some studies move on to evaluate the quality of an action using quantitative kinematic features to predict the development of chronic diseases \cite{ricotti2023wearable,kadirvelu2023wearable,schalkamp2023wearable}. They monitor the execution of exercises by calculating metrics (e.g., speed, range of movement, variations) against \textit{gold standards} \cite{su2014kinect,hoang2016onebody,clarke2020reactive,jo2023flowar}, and detecting the presence of abnormal movement behavior \cite{wang2021leveraging} to estimate the difficulty one may have in specific activities. 
    
    Aside from the studies above on supporting people's healthcare, assistive and fitness needs with diverse ubiquitous technologies, we also found the following studies that move deeper into action understanding, focusing on tasks that require more specific insights or feedback. Guo et al. \cite{guo2018device} highlights the importance of reminding the user to maintain proper posture during workouts. However, the feedback provided in their system is based on the comparison of low-level features like repetition consistency and temporal differences between normal and expert users, which is indirect for this purpose. 
    
    Wei et al. \cite{wei2019towards} proposed a system aimed at facilitating remote training for individuals with Parkinson's disease. This system is capable of action understanding and feedback generation for three actions (referred to as \textit{tasks} in their study), namely squat, forward lunge, and backward lunge. By analyzing angular features computed from the pose estimation by Kinect \cite{Kinect}, the system can identify errors in execution (i.e., violation of criteria set by physios such as \textit{`keep the back knee straight'}). Based on this analysis, the system then provides recommendations, deciding whether the user should repeat the exercise, adjust the difficulty level (for instance, by reducing the rotation angle or introducing additional support), or progress to the next level. In a separate study, Wang et al. \cite{wang2023physiq} only used a smartwatch to evaluate the quality of three upper-arm exercises, namely shoulder abduction, external rotation, and forward flexion, and provide feedback on the number of repetitions, range of movements, and stability of each exercise. Lee et al. \cite{lee2021human} proposed a more comprehensive set of features to measure the performance of stroke patients performing three upper limb exercises. By facilitating a collaborative approach between the AI system and physiotherapists in determining which features to examine, they showed improved evaluation results. These studies rely on what we can categorize as \textit{rule-based} approaches, where the features designed for action evaluation are derived from a convergence of quantitative computational methods (informed by full-body pose or data from a wearable device attached to a specific body part) and qualitative analysis of the specific action by domain experts. A major disadvantage of these approaches is the lack of generalizability of the features, particularly when confronted with more complex movements with diverse patterns \cite{liao2020review}. Also, the feedback provided in these systems mainly evaluates the outcome or quality of the action, and falls short of guiding a person to act more properly.

\subsection{Modeling Actions with Natural Language Description}
    
    Modeling actions with natural language description is essential for understanding and interpreting the semantics of human actions. Early investigations \cite{mtmap1, mtmap2,feature1,feature2,wang2003recent} on such a modeling process mainly focused on human actions with large body movements (e.g., waving the hand), general action categories (e.g., walking around) and explicit descriptions (e.g., counts of a specific gesture pattern). The user's 3D pose series, represented in the positions and attitudes of a set of body joints (e.g., 24 body key points under the SMPL model \cite{10.1145/3596711.3596800}), was taken as the main input of motion-language models. Although Plappert et al. \cite{mtmap1} and Yamada et al. \cite{mtmap2} demonstrated the feasibility of establishing a bidirectional mapping between human whole-body action and natural language, the target pose sequences were usually deterministic or with limited variety, probably due to insufficient action feature descriptions and the hard-supervised training process. To enrich the description space for capturing the dynamics and kinetic features of human motion, Holden et al. \cite{feature1,feature2} utilized rigid body dynamic features such as joint velocity, orientation, and foot contact state for better modeling of human action in temporal and spatial domains. Guo et al. \cite{T2M} introduced a temporal variational autoencoder framework, guided by a learned distribution function, to construct a probabilistic mapping between action sequences and textual descriptions.

    Although the abovementioned work managed to bridge general motions and semantics, the problem become far more challenging given a finer granularity (e.g., mining certain micro-expressions) or specific attention patterns (e.g., telling the difference of different walking patterns) for the description target \cite{VQVAE,zhang2023motiongpt}. This challenge originates from the inherent modality discrepancy between text and action series, which would result in substantial ambiguity and information loss in the interpretation process. From a higher perspective, human motions are far more complex than they appear to be. For example, a ``walking'' action series could potentially indicate the subject's mental state (e.g., casually or hurriedly) and physical state (e.g., any disease-related features from clinical observations). Aiming at these challenges, previous work have investigated different methods and models for enriching action representation in different levels of hidden semantic spaces. For example, VQ-VAE (Vector Quantized Variational Autoencoder) \cite{VQVAE} presented an elegant solution of discretizing the hidden distributions into parameterized ``codebooks'' to increase the capacity of hidden semantics. Guo et al. \cite{TM2T} proposed the use of compact and discrete action tokens to offer a more flexible and adaptable array of action representation for generating text descriptions and the inverse process based on VQ-VAE \cite{VQVAE}, which achieved modeling explicit connections between atomic actions and fine-grained semantics. Kim et al. \cite{kim2022learning} present a joint representation model of human action and language with contrastive learning by leveraging both unpaired and paired action and language datasets. HUMANISE \cite{NEURIPS2022_6030db51} investigated the representation of scene-aware and goal-oriented human motions. Taking advantages of the success of large language models (LLMs), Zhang et al. \cite{T2M-GPT} introduced a LLM-based framework (e.g., a transformer architecture with 12 layers), leveraging the semantic expressiveness of LLMs, to yield high-quality discrete representations of human action given language instructions as input. Later on, other language models like T5 \cite{2020t5,chung2022scaling} and Llama \cite{touvron2023llama} are also adopted to promote the accuracy in building action-language interactions with a similar pipeline \cite{jiang2023motiongpt,zhang2023motiongpt}. In our work, we took the first step to improve the capability of VQ-VAE for representing fine-grained movement patterns and fine-tune language models for comprehensive action description. Therein, we collect feedbacks and comments from clinical physios, so that to provide accurate and professional training guidance.

\section{Method}

    This section details our proposed framework, providing a comprehensive roadmap from the initial preprocessing of action data input to the final generation of detailed feedback. 

\subsection{Biomechanical Features-Driven Action Tokenization}

    In our proposed framework, we begin by discretizing the continuous action data into succinct tokens. This process facilitates the learning of the language model by establishing a mapping between these discrete action tokens and corresponding language tokens.

\subsubsection{Biomechanical Features}
    
    We build on the discretization process driven by low-level features extracted from comprehensive full-body action data. As demonstrated in prior research \cite{TM2T, T2M-GPT, zhang2023motiongpt, jiang2023motiongpt}, this mainly includes each joint's rotation-invariant position, orientation, and velocity. Given a skeleton of 24 joints, as shown in Fig~\ref{fig:fig2}, a total of 287 features can be extracted. Such a methodology enables the generation of descriptive phrases that characterize a person's actions, for example: ``\textit{The person is standing up from a chair and walks forward in a circle}''. 
    
    However, our preliminary experiments, which replicated their pipelines, revealed limitations in producing accurate and consistent descriptions of diverse movement patterns, e.g., ``\textit{This individual is practicing body side bend, but with reduced mobility and a tendency to bend the spine and knees.}'' To help the network learn a more informative discrete codebook and increase its robustness to noise and variations in lower-level features, we have developed an additional set of higher-level features. This was done in collaboration with physios: we transformed the bodily evidences they rely on in their clinical practices into computable biomechanical features. Details of these additional features are as follows: 
    
\begin{figure}[t]
    \begin{minipage}[b]{.3\linewidth}
        \centering
        \includegraphics[width=0.7\linewidth]{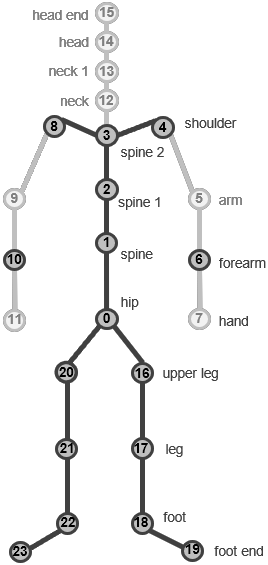}
        \caption{The skeleton of the default action data used in this study. Among the 24 joints, darker ones stand for those used for the computation of biomechanical features.} \label{fig:fig2}
    \end{minipage}%
    \hspace{0.5cm}
    \begin{minipage}[b]{.65\linewidth}
        \centering
        \includegraphics[width=\linewidth]{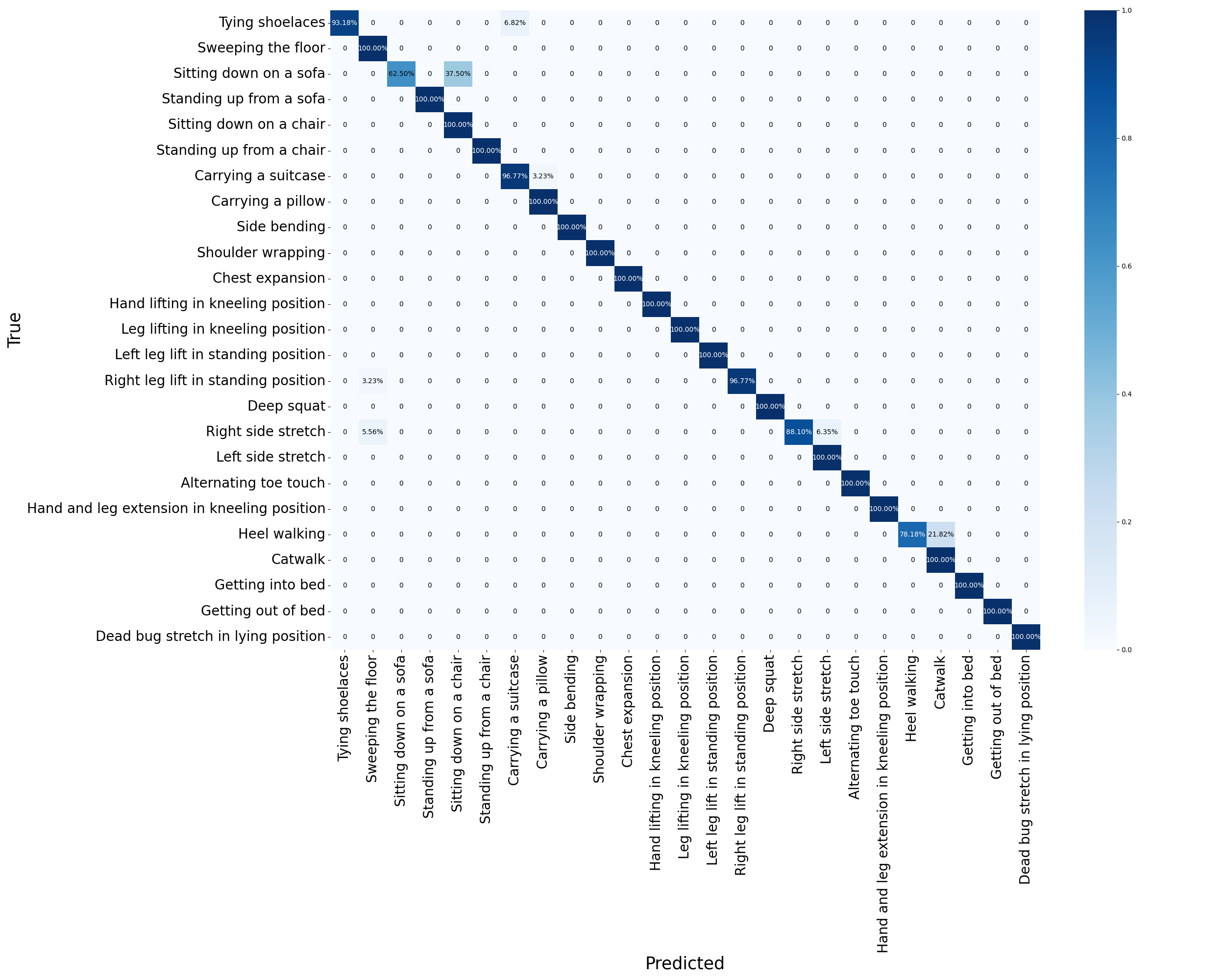}
        \caption{The confusion matrix for the action classification result of our action recognition module.}\label{fig:fig3}
    \end{minipage}
\end{figure}

\noindent$\bullet$ \textbf{Bilateral balance}:
\begin{equation}
    \begin{aligned}
    \mathbf{feats}_{\textit{BB, upper}} &= \mathcal{D}(\mathbf{J}_{\textit{left forearm}}, \mathbf{J}_{\textit{spine2}}) - \mathcal{D}(\mathbf{J}_{\textit{right forearm}}, \mathbf{J}_{\textit{spine2}})\\
    \mathbf{feats}_{\textit{BB, lower}} &= \mathcal{D}(\mathbf{J}_{\textit{left foot}}, \mathbf{J}_{\textit{hip}}) - \mathcal{D}(\mathbf{J}_{\textit{right foot}}, \mathbf{J}_{\textit{hip}}) + \mathcal{D}(\mathbf{J}_{\textit{left leg}}, \mathbf{J}_{\textit{hip}}) - \mathcal{D}(\mathbf{J}_{\textit{right leg}}, \mathbf{J}_{\textit{hip}})\\
    \end{aligned}
\end{equation}

\noindent$\bullet$ \textbf{Inter-joint angle}: 
\begin{equation}
    \begin{aligned}
    \mathbf{feats}_{\textit{IJA}} = [&\mathbf{\angle}(\mathbf{J}_{\textit{left shoulder}}, \mathbf{J}_{\textit{spine2}}, \mathbf{J}_{\textit{spine1}}), \mathbf{\angle}(\mathbf{J}_{\textit{right shoulder}}, \mathbf{J}_{\textit{spine2}}, \mathbf{J}_{\textit{spine1}}), \mathbf{\angle}(\mathbf{J}_{\textit{spine2}}, \mathbf{J}_{\textit{spine1}}, \mathbf{J}_{\textit{spine}}), \\
     &\mathbf{\angle}(\mathbf{J}_{\textit{spine1}}, \mathbf{J}_{\textit{spine}}, \mathbf{J}_{\textit{hip}}), \mathbf{\angle}(\mathbf{J}_{\textit{spine1}}, \mathbf{J}_{\textit{hip}}, \mathbf{J}_{\textit{left upperleg}}), \mathbf{\angle}(\mathbf{J}_{\textit{spine1}}, \mathbf{J}_{\textit{hip}}, \mathbf{J}_{\textit{right upperleg}}),\\
     &\mathbf{\angle}(\mathbf{J}_{\textit{spine1}}, \mathbf{J}_{\textit{hip}}, \mathbf{J}_{\textit{left leg}}), \mathbf{\angle}(\mathbf{J}_{\textit{spine1}}, \mathbf{J}_{\textit{hip}}, \mathbf{J}_{\textit{right leg}}), \mathbf{\angle}(\mathbf{J}_{\textit{left upperleg}}, \mathbf{J}_{\textit{left leg}}, \mathbf{J}_{\textit{left foot}}),\\
     &\mathbf{\angle}(\mathbf{J}_{\textit{right upperleg}}, \mathbf{J}_{\textit{right leg}}, \mathbf{J}_{\textit{right foot}}), \mathbf{\angle}(\mathbf{J}_{\textit{left shoulder}}, \mathbf{J}_{\textit{spine2}}, \mathbf{J}_{\textit{hip}}), \mathbf{\angle}(\mathbf{J}_{\textit{right shoulder}}, \mathbf{J}_{\textit{spine2}}, \mathbf{J}_{\textit{hip}}), \\
     &\mathbf{\angle}(\mathbf{J}_{\textit{left shoulder}}, \mathbf{J}_{\textit{hip}}, \mathbf{J}_{\textit{right shoulder}}), \mathbf{\angle}(\mathbf{J}_{\textit{left leg}}, \mathbf{J}_{\textit{left foot}}, \mathbf{J}_{\textit{left footend}}), \mathbf{\angle}(\mathbf{J}_{\textit{right leg}}, \mathbf{J}_{\textit{right foot}}, \mathbf{J}_{\textit{right footend}})]\\
    \end{aligned}
\end{equation}

\noindent$\bullet$ \textbf{Inter-line angle}: 
\begin{equation}
    \mathbf{feats}_{\textit{ILA}} = \mathbf{\angle}(\mathbf{J}_{\textit{left shoulder}}, \mathbf{J}_{\textit{right shoulder}}, \mathbf{J}_{\textit{left upperleg}}, \mathbf{J}_{\textit{right upperleg}})
\end{equation}

\noindent$\bullet$ \textbf{Inter-joint distance and ratio}:
\begin{equation}
    \begin{aligned}
    \mathbf{feats}_{\textit{IJD}} = [&\mathcal{D}(\mathbf{J}_{\textit{left shoulder}}, \mathbf{J}_{\textit{hip}}), \mathcal{D}(\mathbf{J}_{\textit{right shoulder}}, \mathbf{J}_{\textit{hip}}), \mathcal{D}(\mathbf{J}_{\textit{hip}}, \mathbf{J}_{\textit{left foot}}), \mathcal{D}(\mathbf{J}_{\textit{hip}}, \mathbf{J}_{\textit{right foot}}),\\
    &\mathcal{D}(\mathbf{J}_{\textit{left forearm}}, \mathbf{J}_{\textit{left leg}}), \mathcal{D}(\mathbf{J}_{\textit{left forearm}}, \mathbf{J}_{\textit{right leg}}), \mathcal{D}(\mathbf{J}_{\textit{right forearm}}, \mathbf{J}_{\textit{left leg}}), \mathcal{D}(\mathbf{J}_{\textit{right forearm}}, \mathbf{J}_{\textit{right leg}})]\\
    \end{aligned}
\end{equation}

\begin{equation}
    \mathbf{feats}_{\textit{IJR}} = [\frac{\mathcal{D}(\mathbf{J}_{\textit{left forearm}}, \mathbf{J}_{\textit{right leg}})}{\mathcal{D}(\mathbf{J}_{\textit{left forearm}}, \mathbf{J}_{\textit{left leg}})}, \frac{\mathcal{D}(\mathbf{J}_{\textit{right forearm}}, \mathbf{J}_{\textit{left leg}})}{\mathcal{D}(\mathbf{J}_{\textit{right forearm}}, \mathbf{J}_{\textit{right leg}})}]\\
\end{equation}

    \noindent where $\mathbf{J}_*$ refers to corresponding joints. $\mathcal{D}(*, *)$ represents the Euclidean distance, taking 3D joint positions as input, with \begin{math}\mathcal{D}(a,b) = \sqrt{(a_x-b_x)^2 + (a_y-b_y)^2 + (a_z-b_z)^2}\end{math}
    . The inter-joint (with three joints) and inter-line (with four joints) angle is computed as 
    \begin{math}\angle(a,b,c) = \arctan \left( \frac{||\vec{ab} \times \vec{bc}||}{\vec{ab} \cdot \vec{bc}} \right)\end{math}
    , and 
    \begin{math}\angle(a,b,c,d) = \arctan \left( \frac{||\vec{ab} \times \vec{cd}||}{\vec{ab} \cdot \vec{cd}} \right)\end{math}
    , respectively. It should be noted that the above features are computed for each timestep, presenting potential opportunities for the real-time deployment of our system. The total number of features computed here is 28. We provide such details here as not only to support the replication of this work, but also to inform researchers working on rule-based applications about the features that could be considered for action pattern analysis.

\subsubsection{Action Tokenization}
    
    By computing the features per timestep, for each action sequence of duration $T$, we have $\mathbf{X}=\{x_t \}_{t=1}^{T}$ and $x_i\in\mathbb{R}^{d_{action}}$, where $d_{action}$ is the dimension of features, which is 315 for a pose data input of 24 joints. A VQ-VAE model \cite{VQVAE} is further used to learn the discrete action tokens, using a codebook of length $K$, denoted as $\mathbf{\mathcal{B}}\{b_k\}_{k=1}^{K}$ and $ b_k\in\mathbb{R}^{d_{code}}$. Similar to other encoder-decoder architecture, the encoder of VQ-VAE first encodes the input action features into latent vectors with 1D convolution as $\mathcal{E}(\mathbf{X})=\mathbf{H}$, with $\mathbf{H}=\{h_n\}_{n=1}^{N},N=T/l$, and $l$ denotes the downsampling rate of the convolutional encoder. Then, the quantization through the codebook $\mathbf{\mathcal{B}}$ is carried out to search for the most close counterpart vector $b_k$ for each encoded latent vector $h_n$ as:
    \begin{equation}
        z_n = \underset{b_k\in\mathbf{\mathcal{B}}}{\mathrm{argmin}}\|h_n-b_k\|_2,
    \label{equa1}
    \end{equation}
    \noindent where the sequence of indexes of all the searched $b_k$ in the codebook $\mathbf{\mathcal{B}}$, denoted as $\mathbf{Q}=[q_1,...q_N]$ with $q_i\in[1,2,...,q_N]$, becomes the sequence of tokens for the input action, which is used to fine-tune a language model as described later. Thereon, the decoder reconstructs the input action data with $\mathbf{X_r}=\mathcal{D}(\mathbf{Z})$. The VQ-VAE is trained in a self-supervised way given the loss computed against the input action data as:
    \begin{equation}
        \mathcal{L}_{vqvae} = \mathcal{L}_{recon}+\|\mathbf{H}-sg[\mathbf{Z}]\|_2+\beta\|sg[\mathbf{H}]-\mathbf{Z}\|_2,
    \label{equa2}
    \end{equation}
    \noindent where the three parts of the total loss stand for the reconstruction loss of the VAE, and the embedding and commitment losses of the quantization process, $sg[\cdot]$ denotes stop-gradient operation, $\beta$ is a weighting hyperparameter. Motivated by the velocity regularization used in \cite{T2M-GPT}, the reconstruction loss is added with the regularization from reconstructing the biomechanical features as:
    \begin{equation}
        \mathcal{L}_{recon} = |\mathbf{X}-\mathbf{X_r}|+\alpha|\mathbf{X}[\mathrm{bio}]-\mathbf{X}[\mathrm{bio}]_r|,
    \label{equa3}
    \end{equation}
    \noindent where $\mathbf{X}[\mathrm{bio}]$ stands for getting out the biomechanical features from the original (or reconstructed) data, $\alpha$ is a balancing hyperparameter. The L1 smooth loss is used in our experiment according to existing works \cite{T2M-GPT,jiang2023motiongpt,zhang2023motiongpt}.

\subsubsection{Human Action Recognition Module}

\label{sec:3.1.3}

    During our experiments, we found it effective to insert the action type information into the instruction for tuning language models to improve its performance on describing the detailed movement patterns. A similar finding is also reported in \cite{wang2021leveraging}, where the inclusion of activity type information was found to be advantageous in improving the detection of specific movement behaviors. Furthermore, the action type information can also be used as a query to extract background knowledge from our pre-defined knowledge base, enhancing the feedback of a language model. 
    
    By utilizing the proposed action features extracted from each action instance as input, we trained a 1D convolutional neural network to classify the different action types, i.e., 
    \begin{math}
    y = \mathrm{ConvNet}(\mathbf{X}), y\in\mathbf{Y}_C
    \end{math}
    , where $y$ is the predicted action label, $C$ is the number of action classes. For the dataset we collected, which is described in detail in the next section, the network is able to achieve an average macro F1 score of 0.9636 with a standard deviation of 0.0067, after five different runs of validation (each run divides participants with 85\% for training, 5\% for validation, and 10\% for testing). The confusion matrix is shown in Fig~\ref{fig:fig3}. The mistakes are mostly about misclassifying sitting on the chair vs. on the sofa, and heel-walking vs. heel-to-toe walking. Please kindly refer to Appendix~\ref{appendix} for details about the convolutional backbones used in this module and the VQ-VAE model.
    
\subsection{Action-Conditioned Description with a Language Model}
\label{tuning}

\begin{figure}[t]
  \centering
  \includegraphics[width=\linewidth]{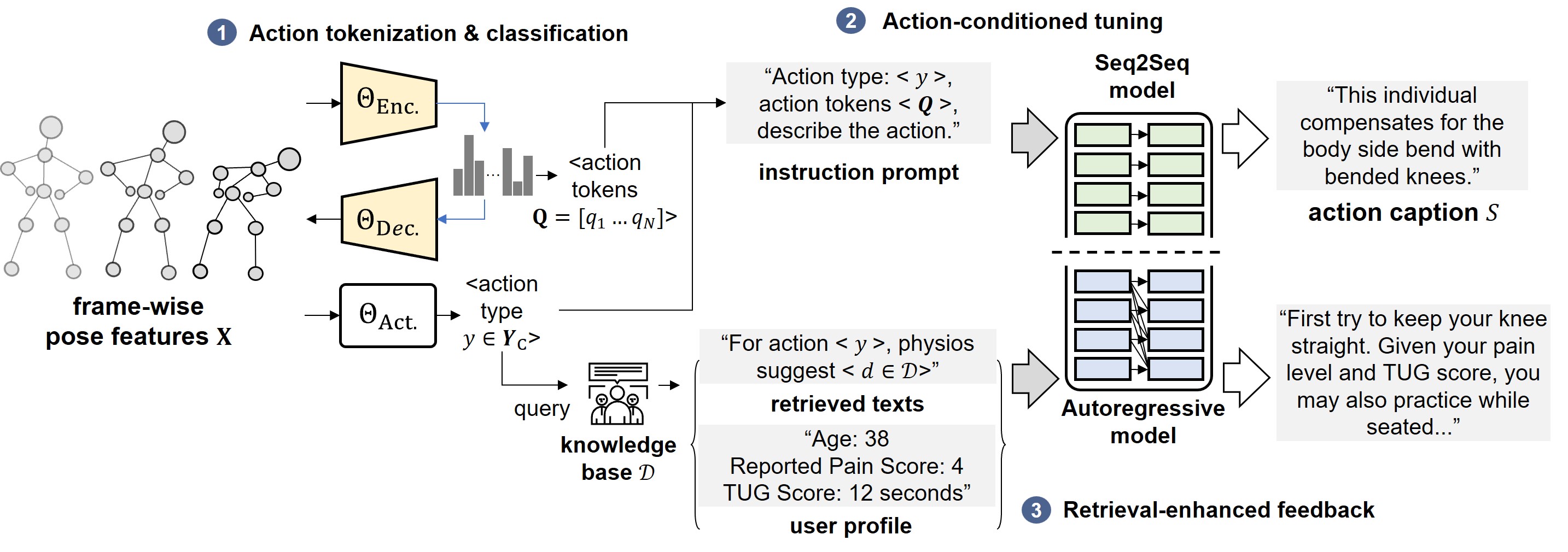}
  \caption{With 3D body poses as input, we first calculate a set of movement features per timestep, which are designed to characterize the local dynamics of each body part and movement patterns. Then, we conduct a self-supervised feature quantization step to acquire discrete tokens for each action instance. We use a Sequence-to-Sequence (Seq2Seq) language model to convert such tokens into descriptive texts, detailing the type of action and potential movement patterns. Thereon, together with the user profile, we further adopt a retrieval-enhanced autoregressive language model to generate physio-like feedback and suggestion.}
  \Description{}
  \label{fig:fig4}
\end{figure}

    An overview of our proposed UbiPhysio framework is shown in Fig~\ref{fig:fig4}. After the step 1 of extracting discrete action tokens, namely the sequence of indexes of codebook representations that found most close to our encoded action features, we conduct the fine-tuning of a language model for action description in step 2. Following the instruction structure adopted in \cite{jiang2023motiongpt,zhang2023motiongpt}, we formalize the instruction prompt for each action instance using a mix of texts and action tokens as follows:

    $\bullet$ Task prompt $\mathcal{T}$ : \{\textit{I want you to act as an action interpreter. Given the type of human action and tokens representing the action, please generate a natural language description of the action.}\}

    $\bullet$ Condition input: \{\textit{The action you need to describe is as following, type:} $[y]$, \textit{tokens:} $[\mathbf{Q}]$.\} 

    The action type information $y$ is particularly added here to make action-conditioned input to the language model, given our experimental results that are reported later. We will also demonstrate that such a piece of information help the model better understand the movement patterns hidden in the action tokens. Thereon, for each action instance, the complete instruction tuning pair we prepare for a language model includes, the complete input prompt $\{\mathcal{T}, y, \mathbf{Q}\}$, and the label of action description $S_{\mathrm{gt}}$. It should be noted that the categorical label $y$ is translated into texts here. The optimization of the language model (LM) is as follows:
    \begin{equation}
        \mathcal{L}_{lm} = -\sum_{\theta}^{L}\log p_\theta(S_{\mathrm{gt}} | \{\mathcal{T}, y, \mathbf{Q}\}),
    \label{equa4}
    \end{equation}
    \noindent where $L$ denotes the output length of the language model. Please kindly refer to Appendix~\ref{appendix} for details about the exact data structure prepared for tuning different language models.
    
\subsection{Retrieval-Enhanced Feedback Generation}
\label{section3.3}
    The common sense knowledge of existing large language models (LLMs) come from their pretraining against a huge amount of textual data. By providing the LLM (taking GPT-4\footnote{The OpenAI ChatGPT (\url{https://chat.openai.com/}) at its August 3rd, 2023 version.} for example) with the detailed action description $S$, it could yet give some generic feedbacks as follows:

    $\bullet$ Prompt: \{\textit{You are acting as an expert of sport rehabilitation, you will support a person (age: 32, reported pain score: 4 (0-10), Time-Up-and-Go score: 12 seconds) during the following session. From now, you will receive detailed action descriptions about their performances, to which you should provide instant feedback to help this person conduct each exercise properly. Action description: [\textit{This individual compensates for the body side bend with \textbf{bended knees}}}].\}

    $\bullet$ Original response: \{\textit{I noticed that you're bending your knees to compensate for the side bend. It's essential to maintain straight knees to get the most benefit from the exercise and to target the right muscles. Focus on \textbf{keeping your knees extended} and try to bend from your waist. This will ensure proper alignment and maximize the benefits of the exercise. Let's try that again and keep those knees straight!}\}

    As is shown, the LLM is able to suggest a direct way of correcting the action and help the user maintain a full engagement of the exercise. However, if such a pattern persists, the language model would continue asking the person to try to stretch the knee, instead of suggesting an option that require less of this person, during our multi-round conversation with it. Whereas, a physio at this moment would suggest, ``\textit{let's try this exercise again while seating on a chair}''. We believe such a piece of knowledge from the domain expert will strengthen the language model to better act vividly and professionally. By providing this information to GPT-4, its response is improved as follows:

    $\bullet$ Knowledge-enhanced response: \{\textit{I understand that you might be facing some challenges with this exercise. To ensure your comfort and safety, \textbf{consider doing the training while seated.} This way, you can get a better grip on the movement without feeling too strained or in pain. We can take it step by step, aiming for gradual progress while staying comfortable.}\}
    
    Here, via a workshop with clinical physios, we collect a rehabilitation knowledge base $\mathcal{D}={d}$ detailing the low-demand and high-demand feedbacks they would consider for different movement patterns per each of the actions that included in this work. The low-demand ones refer to the suggestion that require less of the participant's capacity, it usually proposes an alternative way to perform the action properly, e.g., carry out side-bend against a wall if trunk rotation or hip lateral shift are found. Whereas, the high-demand suggestion point to the way of correcting the movement patterns in a straightforward manner. We organize the data along an index corresponding to different action types. Thus, the knowledge for action type $y$ is retrieved as $d\in\mathcal{D}:\mathrm{I}(d)=y$, where $\mathrm{I}(\cdot)$ denotes the index function.

\section{Experiment}

    This section presents the experimental details, including the collection of a large-scale action-language dataset and the implementation details of our framework. 

\subsection{Data Collection in a Home-Like Space}

\begin{figure}[ht]
  \centering
  \includegraphics[width=0.8\linewidth]{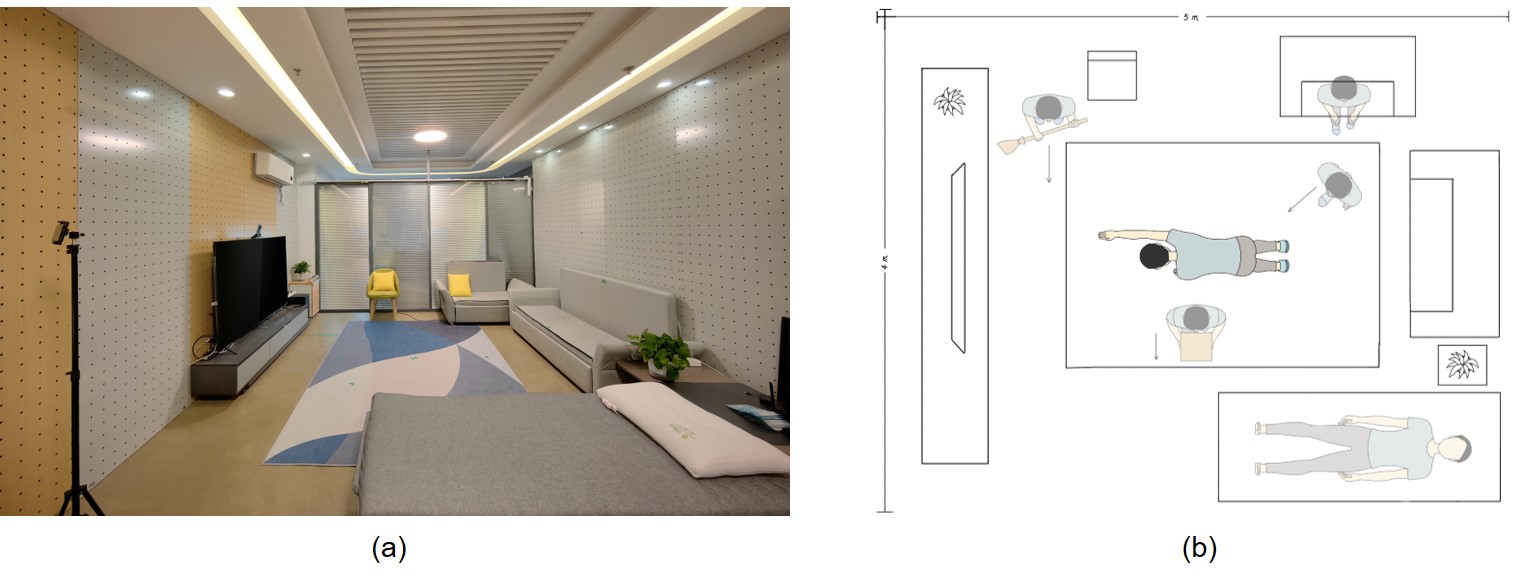}
  \caption{(a) The furnished, home-like space used during data collection, which is prepared to reduce the gap between the lab and the normal living space of a user. (b) The spatial distribution of actions performed in the experiment.}
  \Description{}
  \label{fig:fig5-8}
\end{figure}

    Targeting the future deployment of our proposed UbiPhysio framework at home, we converted a lab room into a modern-style living space, as shown in Fig~\ref{fig:fig5-8}. This space is equipped with essential furniture and equipment that facilitate daily activities and home exercises. These include a television (TV), chair, sofa, soft carpet, bed, cushions, and green plants. Although not visible in the figure, other items such as a broom, yoga pad, and a suitcase loaded with 2.5Kg are also used during the experimental sessions, particularly for daily activities like sweeping the floor and carrying heavy objects. Instead of using stand-alone cameras that may cause discomfort to the participant, we choose to use three mobile phones to record the session, which are small and user-friendly. They are put at the top of the TV (facing the participant at 0\degree), the corner of the space (at 65\degree), and the middle-left border of the space (at 90\degree, where Fig~\ref{fig:fig5-8} (a) is captured). The visual data is collected to aid the annotation from physios and open opportunities for vision-based pose estimation that may speed up the deployment of our framework. This experiment is approved by the Institutional Review Board (IRB) of the University.

\subsubsection{IMUs and the Custom-Made Suit}

\begin{figure}[hb]
  \centering
  \includegraphics[width=\linewidth]{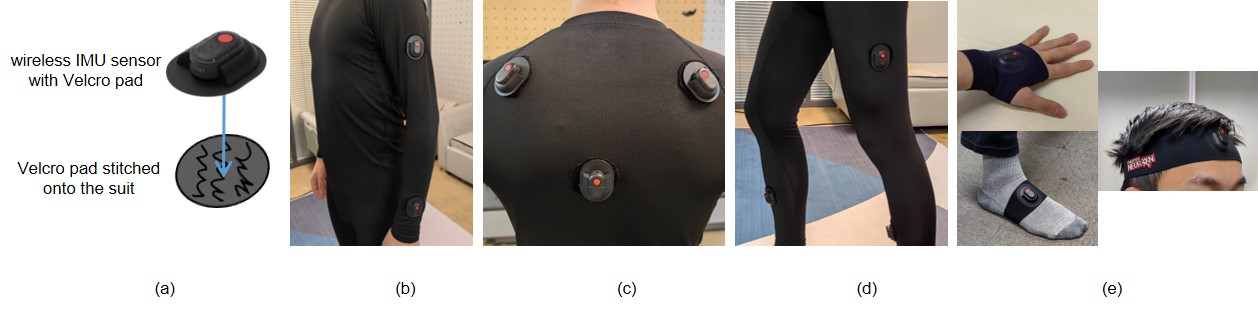}
  \caption{The wearable motion capture device with IMUs. (a)-(d): By stitching Velcro pads onto the anatomical points of our custom-made suit, the wireless IMU sensors with Velcro pads are directly attached onto the suit; this design largely reduces the discomfort of the user, and helps them perform more naturally during the experiment. (e): We additionally used a pair of soft, breathable gloves and three bandages to attach IMUs to the hands, feet, and head respectively.}
  \Description{}
  \label{fig:fig6}
\end{figure}

    To collect the action data, we use the wireless IMU sensors from Noitom \cite{noitom} together with our custom-made suit (as shown in Fig~\ref{fig:fig6}) that allows direct attachment of the sensor without using extensive bandages. The raw IMU data collected by the sensor was processed by manufacturer-provided software to compute pose in 3D coordinates, which constituted the \textit{raw} data utilized in this study. The sensor operates at 60Hz and is equipped with an internal battery and a 2.4G Hz Wi-Fi communication module. Based on our experiences, it typically lasts for a continuous usage of approximately four hours. The suit is available in different sizes (e.g., S, M, L, XL, XXL) and is thin, breathable, and stretchable. It received positive feedback from our participants, with comments such as ``\textit{It wears just like my yoga suit}''. To maintain hygiene, we prepared three suits for each size, ensuring that each participant wore a clean suit at their arrival. It should be noted that for a comprehensive full-body motion capture, we additionally utilized a pair of soft, breathable gloves and three bandages to attach IMUs to the participants' hands, feet, and head, respectively. Totally, 17 IMUs were used, with 12 attached directly onto the suit.

\subsubsection{Daily Activities, Exercises, and Diverse Participants}

\begin{figure}[t]
  \centering
  \includegraphics[width=0.9\linewidth]{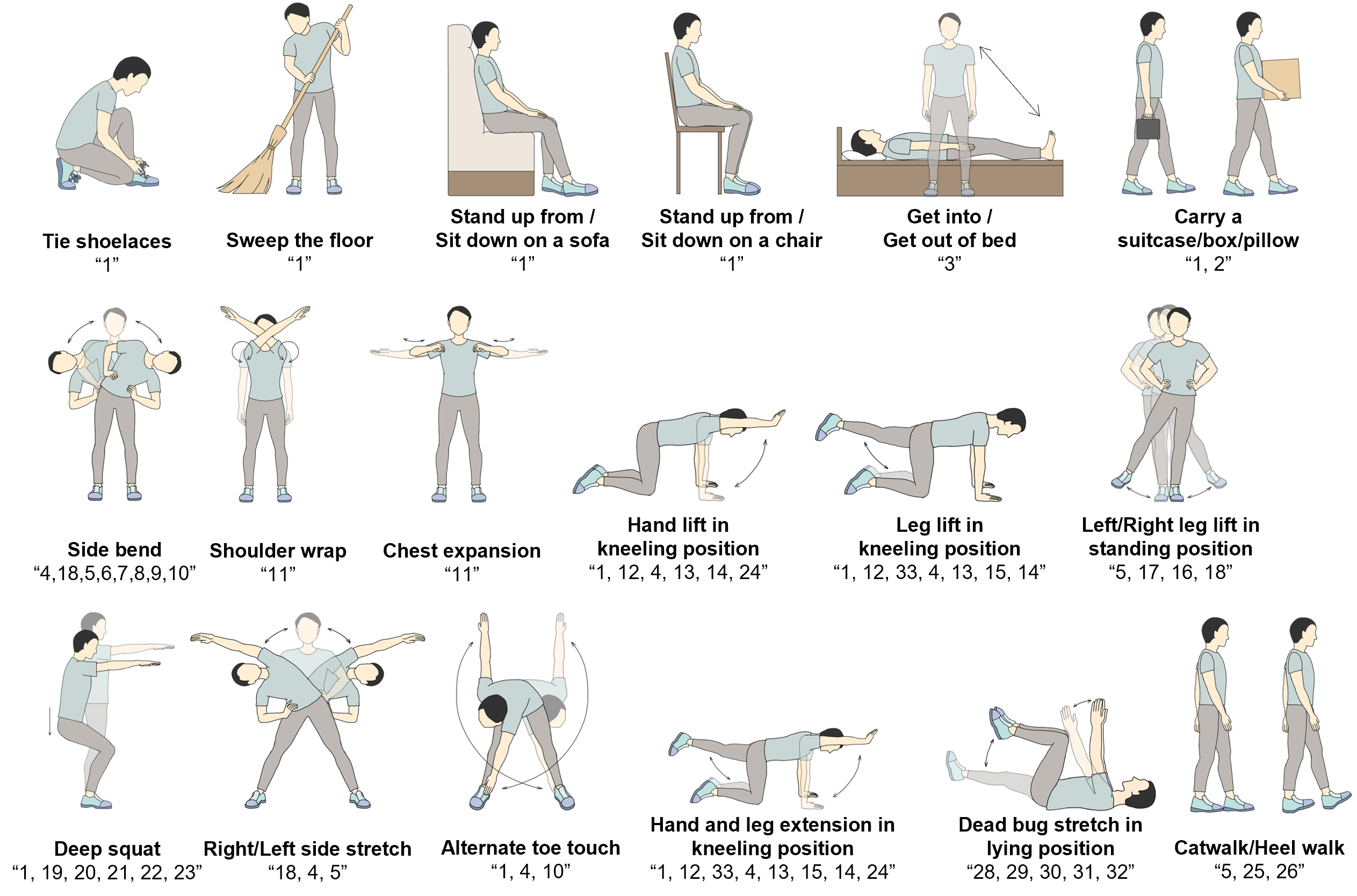}
  \caption{Our experiment considers actions that cover daily functioning, fitness, and rehabilitation activities. Numbers in quotations represent the possible movement patterns, each of which may exist alone or in a combination with others.}
  \Description{}
  \label{fig:fig7}
\end{figure}

    The collected data could be categorized into two groups, namely activities of daily life and exercises that are commonly incorporated in fitness and rehabilitation programs. A comprehensive overview of all actions is shown in Fig~\ref{fig:fig7}. These specific actions were chosen for their effectiveness in engaging the back muscle groups \cite{gorji2022pain,kim2020core,halliday2019randomized} Additionally, our physio collaborators confirmed the safety of these actions for our participants. During recruitment, we set a clear inclusion criteria to exclude individuals with significant health risks or conditions that could impact motion. By advertising on social media and relevant platforms, we recruited 43 healthy participants (22 males, 21 females, with an average age of 28.125 (std: 8.951)) to simulate daily fitness scenarios, and 61 participants with chronic lower-back pain (20 males, 41 females, with an average age of 32.582 (std: 10.433)), i.e., having experienced painful events for at least 3 months, for home-based rehabilitation scenarios. The only difference in our system in dealing with these two groups of participants lies in the generation of feedback from different perspectives. That is, the language model acts as a coach for healthy individuals and as a physio for chronic pain sufferers.

\subsubsection{A Continuous Session with Small Interventions}

    Each participant took a single data collection session, consisting of two stages. First, as we assisted the participant in wearing the suit, we briefed them on the key aspects of the experiment. This includes the designated area for floor sweeping, the suitcase and pillow to be carried, and the path along which walking actions should be performed. We also allowed participants to familiarize themselves with all the actions, offering practice if necessary. Fig~\ref{fig:fig5-8} (b) illustrates the spatial distribution about how these actions are conducted within the experimental space. Secondly, participants performed each action independently. A visual-acoustic reminder was displayed on the television, indicating the action to be performed and counting the remaining repetitions. Although the counting was manually operated by the experimenter, participants were informed that \textit{a real-time automatic action recognition system is in use}. This Wizard-of-Oz design aimed to minimize the experimenter's intervention in participants' behaviors. Except for obvious errors, participants were assured that the experimenter would not intervene, and there was no need to seek confirmation of correctness. Only one experimenter remains in the room during the formal data collection stage. Each action was followed by a rest period, and participants could request longer intervals if needed. The average duration of a complete data collection session was approximately 25 minutes. The annotation of action types was conducted concurrently with each experimental session, where the experimenter marks the onset and offset timestamps for each performed action. A post-hoc inspection is carried out to correct potential errors.
    
    For this study, we left out the transitions between different actions, primarily comprising standing still, casual walking, and self-massage. By extracting each single action instance (i.e., a single execution) and segmenting the longer actions (i.e., sweeping the floor and chest fly) into 15-seconds non-overlapping windows, we obtain 9548 action instances from 104 participants. Each instance lasts from 2 to 20 seconds. While we initially recruited 119 participants, 15 were excluded given IMU-drifting flaws in poses after visual inspection of the data.

\subsection{Collaborating with Clinical Physios}

\renewcommand{\arraystretch}{1.0}
\begin{table}[b]
  \caption{The reference table for the movement patterns included in this work.}
  \label{tab:table1}
  \resizebox{\linewidth}{!}{%
  \begin{tabular}{cp{7cm}|cp{7cm}|cp{7cm}}
    \toprule
    \Large \textbf{Index}      & \Large \textbf{Patterns}       & \Large \textbf{Index}     & \Large \textbf{Patterns}  & \Large \textbf{Index}     & \Large \textbf{Patterns}\\
    \midrule
    1   & \Large Lumbar Flexion/Extension & 12       & \Large Lumbar Hyperextension/Swayback       & 23       & \Large Uneven Bilateral Loading          \\
    2   & \Large Lumbar Hyperextension after Lifting Heavy Objects & 13       & \Large Trunk Deviation from Midline/Trunk Lateral Shift       & 24       & \Large Thoracic Hyperextension Compensating for Shoulder Joint Movement          \\
    3   & \Large Getting Up Directly from / Lying Down Directly into a Supine Position & 14       & \Large Hip Hyperflexion       & 25       & \Large Insufficient Ankle Dorsiflexion           \\
    4   & \Large Trunk Rotation            & 15       & \Large Upper Chest Depression       & 26       & \Large Trunk Anterior Lean
          \\
    5   & \Large Hip Lateral Shift           & 16       & \Large Hip Tilt       & 27       & \Large Incorrect Walking Pattern           \\
    6   & \Large Spinal Extension           & 17       & \Large Lumbar Lateral Flexion      & 28       & \Large Lumbar Lift off the Bed           \\
    7   & \Large Cervical Lateral Flexion Compensation            & 18       & \Large Trunk Flexion       & 29       & \Large Same-side Hand and Foot Movement           \\
    8   & \Large On tiptoes/Pelvic Tilt           & 19       & \Large Excessive Anterior Knee Displacement       & 30       & \Large Head Not Touching the Ground           \\
    9   & \Large No Trunk Activity           & 20       & \Large Upright Trunk Squat       & 31       & \Large Thigh Not Perpendicular to the Floor           \\
    10  & \Large Knee Flexion Compensation           & 21       & \Large Excessive Hip and Knee Flexion Angle/Too Deep Squat       & 32       & \Large Calf Not Parallel to the Bed Surface/Calf Dangling           \\
    11  & \Large Lumbar Hyperextension/Pelvic Anterior Tilt           & 22       & \Large Shallow Squat       & 33       & \Large Hip hyperextension leading to lumbar hyperextension          \\
  \bottomrule
\end{tabular}%
}
\end{table}
\renewcommand{\arraystretch}{1.0}

    A committee consisting of 3 clinical physios was established to aid this work. The physios were recruited from the rehabilitation center of a national hospital. They first helped design inclusion criteria for participants, action types to be considered, and the procedure of sessions. Their other tasks include annotating possible movement patterns (as detailed below), giving practical feedback on diverse action performances via a workshop (as part of the method described in Section~\ref{section3.3}), and evaluating the language output of our proposed framework (as reported in results of the user study in Section~\ref{section5.2}). 
    
    As shown in Table~\ref{tab:table1}, we first settled down the scope of movement patterns that physios normally consider in their clinical practice, the indexes of these patterns are also provided for each action depicted in Fig~\ref{fig:fig7}. These movement patterns typically serve as evidence for physios to provide feedback and support during clinical rehabilitation sessions, as well as in telemedicine scenarios \cite{biely2014clinical}. Compared to the metrics used in some previous studies \cite{wang2021leveraging, wang2019recurrent,wang2021chronic}, they offer more insight into the specific improper movement behaviors of different body parts. Furthermore, this extensive set of patterns suggests that traditional machine learning methods might require downsampling these patterns into fewer discrete classes to maintain effective performance. In contrast, modeling the action patterns using natural language presents as a more flexible and informative approach.
    
    By accessing synchronized video data collected from three different viewing angles during the experiment, physios annotated each individual action instance. Particularly, they selected one or more patterns from the candidates reported in Table~\ref{tab:table1}. Therefore, it was possible for multiple movement patterns to be marked at the same time for a single action instance. Throughout this annotation process, we held discussions to reach agreements among physios. Based on the annotations, we employ GPT-4 to generate the natural language description for each action type with different movement patterns. To guide this process, we provided templates outlining the essential elements of the descriptions, ensuring they encompass both the action type and the associated movement patterns. Three different descriptions are generated for each action-pattern combination. The average length of the description is 15 words, with a maximum of 25 words. Examples of the description are as follows:

    $\bullet$ Action type: <shoulder wrap>, movement patterns: <11-lumbar hyperextension>, description: ``\textit{When executing a shoulder wrap, this individual exhibits poor core stability and their waist is overextended.}''.
    
    $\bullet$ Action type: <hand lift in kneeling position>, movement patterns: <4-trunk rotation, 13-trunk lateral shift>, output: ``\textit{In the kneeling hand forward exercise, this individual moves their trunk laterally.}''.
    
    $\bullet$ Action type: <leg lift in kneeling position>, movement patterns: <12-Lumbar hyperextension, 14-Hip hyperextension>, output: ``\textit{This individual overextends their hip, causing their back to arch excessively when performing the kneeling leg backward exercise.}''.
    
\subsection{Implementations}
\label{implementation}

    During the feature extraction process, the raw 3D pose data is normalized through skeleton normalization and rotation to face the Z+ direction. After feature extraction, the dimensionality of the action data is 315. The first 287 features are derived from methodologies used in previous studies \cite{TM2T,T2M-GPT,jiang2023motiongpt,zhang2023motiongpt} and the last 28 features are the ones proposed as above in this work to help capture the more subtle movement patterns. For all experiments, the data collected from the 104 participants was split into training, validation, and test sets, comprising 85\%, 5\%, and 10\% of the data, respectively. It was ensured that there was no subject overlap among these sets. We report the metrics scores obtained on the test set for each method. We additionally report the 95\% confidence interval to illustrate the variability in performance, using results from different participants or folds. For language modeling, we evaluate various open-sourced foundational language models, including T5 (in its small, base, large, and 3B variants) \cite{2020t5}, Llama 2 7B\cite{touvron2023llama}, and ChatGLM2 6B\cite{du2022glm,zeng2022glm}. We use an extensive set of benchmark metrics to evaluate the output of each language model against the annotated description, including Bleu \cite{papineni2002bleu}, Rouge (F-scores) \cite{lin2004rouge}, Cider \cite{vedantam2015cider}, and BertScore (F-scores) \cite{zhang2019bertscore}. To reveal significance in the pair-wise comparison, Wilcoxon Signed-rank test is used. For multiple comparisons, Friedman test is additionally used. These statistical tests are conducted with per-metric and per-subject scores.
    
    The duration of a complete training on a single RTX 4090 graphics card of VQ-VAE is 18 hrs, and are as follows for the action-language modeling: T5 small (3 hrs), T5 base (4.9 hrs), T5 large (4.5 hrs), T5 3B (8.9 hrs), Llama 2 7B (5 hrs, for 3 epochs), and ChatGLM2 6B (24 hrs). It should be noted that, for smaller models, the training could have met the GPU inefficiency problem, leading to suboptimal utilization of the high-performance graphics card and longer training time. Please kindly refer to Appendix~\ref{appendix} for details of the hyperparameters used for each model, and the computation of different metrics.

\subsection{User Study}

\begin{figure}[ht]
    \centering
    \includegraphics[width=0.3\linewidth]{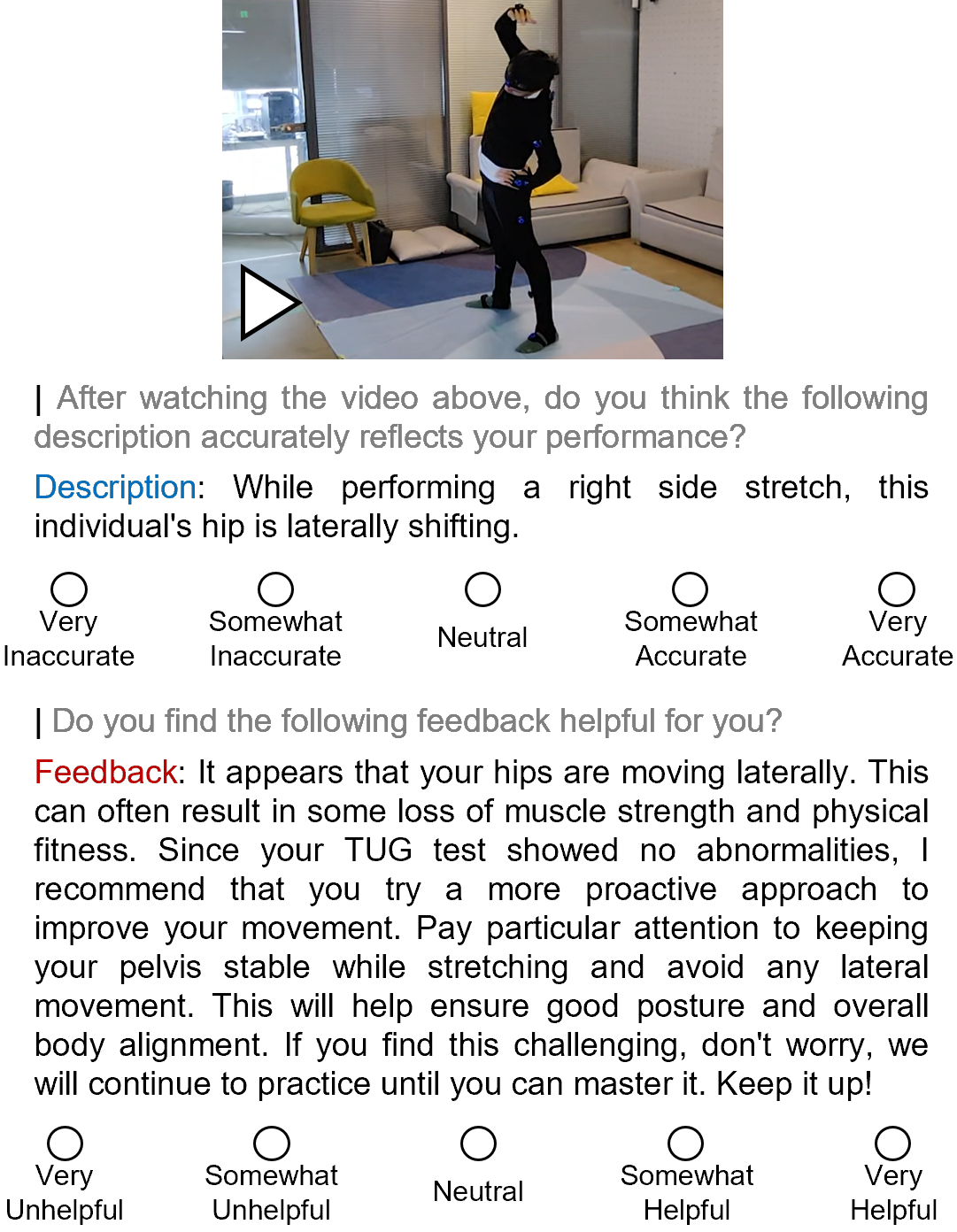}
    \caption{The adopted questionnaire, rated by the participant as shown in the video and physios.}
    \Description{}
    \label{fig:fig9}
\end{figure}

    Aside from using standard objective benchmarks to evaluate outputs of the language model, we also present a straightforward picture of the performance of our proposed framework by collecting feedback from prospective users and physios. We conduct a user study with these groups to evaluate the quality and acceptability of the action description and feedback generated by the language model. For action description, the T5 large model with a window length of 64 and action-conditioned instruction tuning is used. For retrieval-enhanced feedback generation, we utilize GPT-4 with the following prompt: ``\textit{You are now a physiotherapist to support the daily functioning and exercise of a person. In the following, you will receive the <knowledge> specifying the pre-defined low-demand and high-demand feedbacks for the action, <user profile> indicating the age, self-reported pain score (0-10), and Timed-Up-and-Go (TUG) score, <action description> detailing the action type and movement patterns of the person. You need to return the <instant feedback> in a vivid tongue}''. For healthy participants, the role of a physiotherapist is transferred to a fitness coach.

    We create a questionnaire that includes both the model-generated outputs and the actual action descriptions (ground truth). Feedbacks generated with or without our retrieved domain knowledge are also included. These elements are randomly organized in the questionnaire presented to users, in a 50\% by 50\% ratio, in order to eliminate any potential bias. The questionnaire, as shown in Fig~\ref{fig:fig9}, is composed of multiple sections, each presenting a video of an action instance, paired with its action description and feedback. We reach out to 15 participants that are randomly selected from the testing folds, and invite them to complete the questionnaire, where each of them rates upon their own videos. In addition to participants, the physio collaborators also participate in this user study from a professional perspective. They rate action description and feedback for each video using a standard 5-point rating scale.

\section{Results}
    
    The objective evaluation of our framework in terms of its language output is reported in this section, where we look at performances given different foundational language models and tuning strategies that proposed in this work. Additionally, we conducted a user study with participants and physios to collect the subjective preference and opinions about the output of our framework. Finally, we look into the practical deployment of our system in real-world scenarios without specialized devices, highlighting the potential and limitations of our framework. 

\subsection{Quantitative Evaluation of Action Description with Language Models}
\label{comparison}

\begin{table}[t]
  \caption{The comparison of different foundational language models on our action description task with different feature sets as input. Params.= the number of trainable parameters, Dim. = the dimension of the input action data, Iter. = the total number of training iterations. The best results under different feature inputs are highlighted in bold.}
  \label{tab:table2}
  \resizebox{\linewidth}{!}{%
  \begin{tabular}{l | llll | rrrrrrr}
    \toprule
    \textbf{Model}     & \textbf{Params.}     & \textbf{Tuning} & \textbf{Dim.} & \textbf{Iter.} & \textbf{Bleu@1} $\uparrow$ & \textbf{Bleu@4} $\uparrow$ & \textbf{Rouge 1} $\uparrow$ & \textbf{Rouge 2} $\uparrow$ & \textbf{Rouge L} $\uparrow$ & \textbf{Cider} $\uparrow$ & \textbf{BertScore} $\uparrow$ \\
    \midrule
    
    T5 small      & 80M  & Full-params. & 287 & 50K & 63.40$\pm$2.79 & 41.94$\pm$4.14 & 58.71$\pm$3.05 & 44.08$\pm$3.95
                                                    & 55.45$\pm$3.2 & 1.44$\pm$0.16 & 56.80$\pm$3.23   \\
    T5 base       & 248M & Full-params. & 287 & 50K & 61.56$\pm$2.78 & 39.04$\pm$3.75 & 57.12$\pm$2.89 & 41.60$\pm$3.76 
                                                    & 53.21$\pm$3.01 & 1.31$\pm$0.15 & 53.47$\pm$3.22   \\
    T5 large      & 783M & Full-params. & 287 & 50K & \textbf{66.38}$\pm$2.21 & \textbf{44.75}$\pm$3.56 & 62.37$\pm$2.45 
                                                    & 48.02$\pm$3.35 & \textbf{60.01}$\pm$2.59 & \textbf{1.49}$\pm$0.14 & \textbf{58.52}$\pm$2.66   \\
    T5 3B         & 3B   &LoRA        & 287 & 100K  & 65.37$\pm$3.32 & 43.29$\pm$4.82 & 62.55$\pm$3.64 & 48.23$\pm$4.78 
                                                    & 59.86$\pm$3.76 & 1.42$\pm$0.2 &  56.04$\pm$3.68  \\
    ChatGLM2 6B   & 6B   & P-tuning v2 & 287 & 10k  & 38.67$\pm$1.11 & 13.56$\pm$1.28 & 35.35$\pm$1.21 & 17.76$\pm$1.35 & 32.93$\pm$1.19 & 0.13$\pm$0.05 & 21.86$\pm$1.04 \\
    \midrule
    
    T5 small   & 80M  & Full-params. & 315 & 50K    & 64.09$\pm$2.55 & 42.89$\pm$3.94 & 60.64$\pm$2.65 & 46.24$\pm$3.73 
                                                    & 57.74$\pm$2.81 & 1.44$\pm$0.16 & 56.70$\pm$2.87   \\
    T5 base    & 248M & Full-params. & 315 & 50K    & 65.09$\pm$2.49 & 42.86$\pm$3.74 & 60.93$\pm$2.62 & 46.61$\pm$3.49 
                                                    & 57.68$\pm$2.83 & 1.43$\pm$0.15 & 56.29$\pm$2.82   \\
    T5 large   & 783M & Full-params. & 315 & 50K    & \textbf{68.21}$\pm$2.45 & \textbf{47.64}$\pm$3.38 & \textbf{64.74}$\pm$2.78 
                                                    & \textbf{50.95}$\pm$3.38 & \textbf{61.39}$\pm$2.85 & \textbf{1.62}$\pm$0.14 & \textbf{59.84}$\pm$2.93   \\
    T5 3B      & 3B   & LoRA        & 315 & 100K    & 66.70$\pm$3.36 & 45.67$\pm$4.9 & 62.64$\pm$3.68 & 49.35$\pm$4.8 
                                                    & 60.47$\pm$3.8 & 1.51$\pm$0.19 & 57.69$\pm$3.85   \\
    ChatGLM2 6B   & 6B   & P-tuning v2  & 315 & 10K & 57.54$\pm$1.45 & 33.78$\pm$1.84 & 53.46$\pm$1.68 & 37.63$\pm$1.95 & 49.90$\pm$1.71 & 1.07$\pm$0.07 & 48.94$\pm$1.63 \\
  \bottomrule
  
\end{tabular}%
}
\end{table}	

    We first compare the performance of open-sourced foundation models on our action description task with benchmark metrics for natural language generation tasks. The action tokens provided here belong to two groups, the first is acquired by training the VQ-VAE with the 287 features that widely adopted in recent action-language modeling studies \cite{TM2T,T2M-GPT,jiang2023motiongpt,zhang2023motiongpt}, while the other is acquired by adding our proposed biomechanical features that provide the input features to VQ-VAE in a dimension of 315. A window length of 64 is applied during the training of VQ-VAE. The action label is removed from the tuning instructions, and the description label comprises full descriptions of the action type and movement patterns. Results of a standalone cross validation experiment are reported in Table~\ref{tab:table2}. 

    The results confirm the effectiveness of our proposed biomechanical features (Dim.=315) in assisting the VQ-VAE model, and consequently the language models, in better capturing the diverse movement patterns hidden behind each action. Except for T5 small, the use of action tokens driven by these features led to improvements across all metrics of different models. The improvements are significant for T5 base and ChatGLM2 6B ($p<0.05$). For T5 small (Bleu@1 ($p=0.048$), Rouge 1 ($p=0.035$), Rouge L ($p=0.022$)), T5 large (Cider ($p=0.01$), Rouge 2 ($p=0.095$), BertScore ($p=0.076$)), and T5 3B (Bleu@4 ($p=0.083$), BertScore ($p=0.055$)), improvements are also significant or approaching significance. It is worth noting that these features are computed per timestep, providing a promising opportunity for real-time operations.

    As the results indicate, the T5-large model generally demonstrates the most promising performance in this action description task, despite its smaller size compared to large language models (LLMs) such as Llama 2 7B and ChatGLM2 6B. The Llama 2 7B model's fine-tuning on our data for this task was unsuccessful, despite our attempts to modify various hyperparameters of the LoRa framework, learning rate, and input/output lengths. In the first section of this table, while the Bleu scores of T5 3B are smaller than T5 large,  the Rouge-N scores of them are the inverse. This could suggest that, since Bleu focuses on the precision alone, T5 3B generated texts that better cover words in the ground truth but either with more inaccurate words or with a shorter length and was penalized. Therein, given the ground truth of ``While executing a right leg lift, this individual is showing hip lateral shift'', the output from T5 3B is ``This individual is raising their right leg while standing'', while T5 large predicts ``This individual lacks trunk stability whilst performing a right leg lift''. Although both of the methods capture 6 words in the ground truth, the Bleu scores of T5 3B were penalized for its shorter length. The Rouge-N scores of T5 3B are higher because this metric does not penalize the shorter length. The better performance of T5 large is measured by Rouge-L, which considers the longest common subsequence instead of fragmented words. Generally, an objective evaluation of language models benefits from a diverse set of metrics. Additionally, incorporating human preferences, though more resource-intensive, offers valuable insights.
    
    In general, the performance disparities among the models may stem from their different architectures and training objectives. The task at hand resembles language translation, converting discrete action tokens (a form of ``second language'') into formal language. This task might be more effectively tackled by sequence-to-sequence (Seq2Seq) models, such as the T5 family. For LLMs like Llama, the ``translation task'' becomes relatively more challenging due to their decoder-only architecture and autoregressive objective. However, ChatGLM2 6B does produce some meaningful results, which may be owed to its blank-filling objective during pre-training, facilitating fine-tuning in a Seq2Seq manner. 
        
\subsubsection{Impact of Window Length}

\begin{table}[t]
  \caption{The ablation experiment on different window lengths applied during action tokenization, sampling rate, and action-conditioned instruction tuning. Action. = the action type information as a condition added to the instruction. The relatively higher performances in each comparison are marked in bold.}
  \label{tab:table3}
  \resizebox{\linewidth}{!}{%
  \begin{tabular}{l | p{1.5cm}p{1.5cm}l | rrrrrrr}
    \toprule
    \textbf{Model}     & \textbf{Window length}  & \textbf{Sampling rate}   & \textbf{Instruction} & \textbf{Bleu@1} $\uparrow$ & \textbf{Bleu@4} $\uparrow$ & \textbf{Rouge 1} $\uparrow$ & \textbf{Rouge 2} $\uparrow$ & \textbf{Rouge L} $\uparrow$ & \textbf{Cider} $\uparrow$ & \textbf{BertScore} $\uparrow$ \\
    \midrule
    T5 large    & 64  & 60Hz & Tokens & 65.78$\pm$1.95 & 46.20$\pm$2.43 & 64.50$\pm$1.81 & 50.12$\pm$2.44 & 62.11$\pm$1.97 & 1.58$\pm$0.09 & 59.22$\pm$2.38 \\
                & 128 & 60Hz & Tokens & 65.63$\pm$0.45 & 45.95$\pm$0.56 & 63.79$\pm$0.53 & 49.69$\pm$0.62 & 61.32$\pm$0.60 & 1.58$\pm$0.03 & 58.95$\pm$0.74 \\
                & 192 & 60Hz & Tokens & \textbf{66.20}$\pm$0.68 & \textbf{46.99}$\pm$1.27 & \textbf{64.96}$\pm$0.64 & \textbf{50.95}$\pm$1.21 & \textbf{62.45}$\pm$0.78 & \textbf{1.62}$\pm$0.07 & \textbf{60.44}$\pm$1.23 \\
    \midrule
    T5 large    & 96  & 30Hz & Tokens & 65.79$\pm$1.12 & 46.20$\pm$1.69 & 64.20$\pm$1.30 & 50.15$\pm$1.77 & 61.99$\pm$1.52 & 1.59$\pm$0.06 & 59.42$\pm$1.28 \\
    \midrule
    T5 large   & 64  & 60Hz & Action.+Tokens & \textbf{68.97}$\pm$0.76 & \textbf{50.06}$\pm$1.10 & \textbf{67.82}$\pm$0.99 & \textbf{54.23}$\pm$1.34 & \textbf{65.75}$\pm$1.06 & \textbf{1.74}$\pm$0.04 & \textbf{62.78}$\pm$0.78 \\
               & 128 & 60Hz & Action.+Tokens & 67.57$\pm$0.89 & 48.37$\pm$1.28 & 66.58$\pm$1.04 & 52.62$\pm$1.30 & 64.23$\pm$1.13 & 1.68$\pm$0.05 & 61.40$\pm$0.71 \\
               & 192 & 60Hz & Action.+Tokens & 68.59$\pm$1.16 & 49.92$\pm$1.46 & 67.74$\pm$1.20 & 53.93$\pm$1.58 & 65.42$\pm$1.25 & 1.73$\pm$0.06 & 62.67$\pm$1.60 \\
    \midrule
    T5 large   & 192  & 60Hz & Action.+Tokens & \textbf{60.25}$\pm$0.89 & \textbf{47.18}$\pm$1.17 & \textbf{60.04}$\pm$1.06 & \textbf{49.51}$\pm$1.22 & \textbf{59.44}$\pm$1.00 & \textbf{1.71}$\pm$0.11 & \textbf{55.50}$\pm$1.02   \\
    (patterns only)           & 192  & 60Hz & Tokens & 59.33$\pm$1.40 & 45.46$\pm$2.02 & 58.90$\pm$1.52 & 47.79$\pm$1.92 & 58.18$\pm$1.53 & 1.63$\pm$0.1 & 55.43$\pm$1.39   \\
  \bottomrule
\end{tabular}%
}
\end{table}
    
    For action tokenization with the VQ-VAE model, a sliding window is used to accommodate action feature inputs of different lengths. To evaluate the impact of this variable window length, we conducted an experiment using the T5 large model. This model was chosen due to its generally better performances with a smaller size when compared to other language models in the last subsection. The experiment was conducted across five different cross-validation folds, and the results are presented as an average performance, accompanied by a 95\% confidence interval. The detailed results are reported in Table~\ref{tab:table3}. A key finding from this experiment is that the performance of language modeling does not appear to be sensitive to the length of the sliding window employed during action tokenization. The Friedman test only reports significance for Rouge 1 ($\mathcal{X}^2=9.042, p=0.011$) and Rouge L ($\mathcal{X}^2=7.042, p=0.03$) scores of three methods in the third section. This could be attributed to the fact that the tokens are extracted from complete action sequences, thereby potentially neutralizing the impact of varying window lengths. In a secondary experiment, we downsampled the raw action data from 60Hz to 30Hz and tested the model with a window length of 96 (equivalent to the 192 used for data at 60Hz). The results showed a slight decrease in performance, suggesting that a higher data frequency might enhance the sufficiency of action descriptions. However, this reduction in performance was not significant across all the metrics ($p>0.05$). Therefore, the decision to lower the frequency could be considered as a viable strategy for reducing hardware costs and computational loads without substantially compromising model performance.

    \subsubsection{Impact of Action-Conditioned Instruction Tuning} As reported in Table~\ref{tab:table3}, integrating the action type information, denoted as $y$, into the instruction significantly enhances the language model's description performance ($p<0.05$). This enhancement is two-fold. Firstly, the model becomes more proficient at describing the correct type of action. Secondly, it exhibits an improved capacity to capture diverse movement patterns. This strategy is particularly beneficial for systems that engage directly with users in a home environment. It serves as a reliability measure for the language model's output, especially considering the high accuracy achieved in classifying different actions using an efficient classifier, as documented in Section~\ref{sec:3.1.3}. To further validate whether this strategy indeed improves the model's ability to capture movement patterns, we conducted an additional experiment. In this experiment, the model was trained exclusively with labels that contained only pattern information. The results, presented in the last two rows of Table~\ref{tab:table3}, confirm our hypothesis. The enhancement brought about by the integration of action type information is not confined to the accurate description of the action type alone. It also extends to the precise depiction of movement patterns, demonstrating that the improvement is comprehensive and not overshadowed by the correct identification of action types.

\subsection{Insights from the User Study}
\label{section5.2}

\begin{figure}[t]
    \centering
    \includegraphics[width=\linewidth]{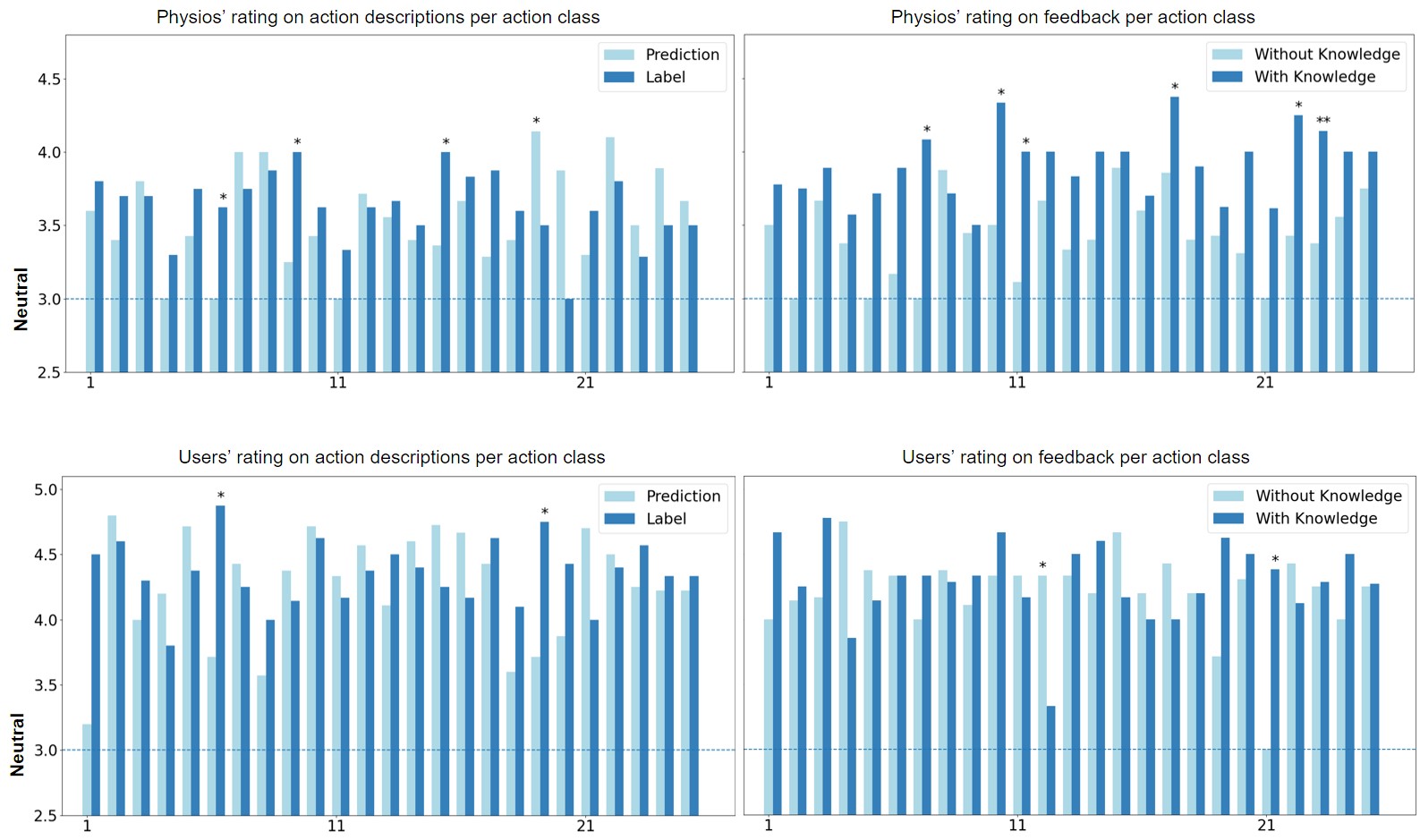}
    \caption{Results of the user study on rating the outputs of our framework per each action type by clinical physios and normal participants. With Wilcoxon signed-rank test on per-subject scores, significant results are marked with * for $\alpha = 0.05$ and ** for $\alpha = 0.01$. Prediction: The action description output by our framework; Label: The ground truth description generated by GPT-4 given annotations from physios. The feedback without knowledge is generated directly by GPT-4 given the prompt comprising user profile and action descriptions, without using the clinical knowledge we gathered from physios, vice versa.}
    \Description{}
    \label{fig:fig10}
\end{figure}
    
    The average results of the study per each group of the participants and physios are reported in Fig~\ref{fig:fig10}. For action descriptions, both participants and physios evaluated the model outputs of most actions as comparable to or even better than the ground truth. However, a subset of the actions was rated with a difference that exceeded the significance level, indicating room for improvement. When describing the action of standing up from a chair (Action \#6), the model consistently generated outputs such as `\textit{this individual is rising from a chair, curving their back forward}'. It appears that distinguishing between normal patterns and improper back curvature is challenging for the model due to their high similarities. In the case of the left-leg lift in a standing position (Action \#15), the model frequently predicted the users' performances as unstable - a description criticized by physiotherapists but accepted by participants, who often found this action difficult to execute while maintaining balance. More interestingly, for the action of alternating toe touch (Action \#19), while physiotherapists agreed with the model's prediction of the performances as `\textit{rounding their back excessively}', some participants believe they performed the action correctly. In a clinic setting, physiotherapists would actively engage with the patient to determine if such a discrepancy is due to a lack of capacity (i.e., the performance is the best the patient can do) or cognitive bias (i.e., the patient is unaware of the harm in performing the action in that particular manner). Based on this interaction, they would come to a consensus and decide whether to adjust the action's difficulty or work on improving the patient's self-awareness. \textbf{We believe such an interaction could also happen between our framework and a user, by further enabling a multi-round dialogue between them that simulates such an in-depth patient-physio interaction}. This may open opportunities for personalized services and improve the model's efficacy via continual learning. We would like to leave this to its future development.

    For generated feedback, there was a noticeable divergence in the assessment provided by the physiotherapists and the participants. The physiotherapists rated the retrieval-enhanced feedbacks, those generated with domain knowledge, as significantly superior to the plain ones generated without such knowledge. This highlights the importance of domain knowledge in improving the quality of generated feedback. By contrast, the participants exhibited a more neutral stance, indicating no strong preference between the two types of feedback. This discrepancy in ratings may due to the lack of professional knowledge in our participants. Consequently, they may perceive the feedback provided by GPT-4, based on its common sense knowledge, as sufficiently satisfactory. For the action of hand lift in a kneeling position (Action \#12), we observed that the feedback generated by the model without domain knowledge tended to be more fluent and natural. When domain knowledge was incorporated, the feedback became somewhat rigid and unnatural. This could explain why participants found the former feedback significantly more helpful.

    In general, the participants and physios are all excited about the achievement made by our proposed framework. During the user study, we also collect some opinions regarding potential improvements of our framework from both participants and physios. Most of them underscored the need for further refinement of the model's language use and tone to enhance the acceptability and effectiveness of the conversation. For example, the model tends to use the language such as `\textit{one is carefully conducting an action}', which is criticized by our physiotherapists due to its ambiguity in action description. Such language might be interpreted in multiple ways. Some may perceive it as indicating that the individual is performing the action slowly yet with balance, while others might interpret it as a sign of the difficulties the person is encountering during the action. Furthermore, they point to the importance of integrating visual demonstrations alongside textual instructions on how to execute certain actions. A physio also highlights the need for the feedback module to be aligned with the user's specific symptoms. This is particularly important given the heterogeneity of chronic lower-back pain, which manifests in varied subtypes, such as the differences in the localized affected regions.
    
\subsection{Preliminary Evaluation towards Real-Life Deployment}

    Here, we provide an initial examination of the feasibility of implementing our framework for practical use in home settings. The considerations primarily revolve around two aspects: the system's inference time, spanning from action data preprocessing to feedback generation, and the extra devices needed to run this framework.

\begin{figure}[b]
    \centering
    \includegraphics[width=0.7\linewidth]{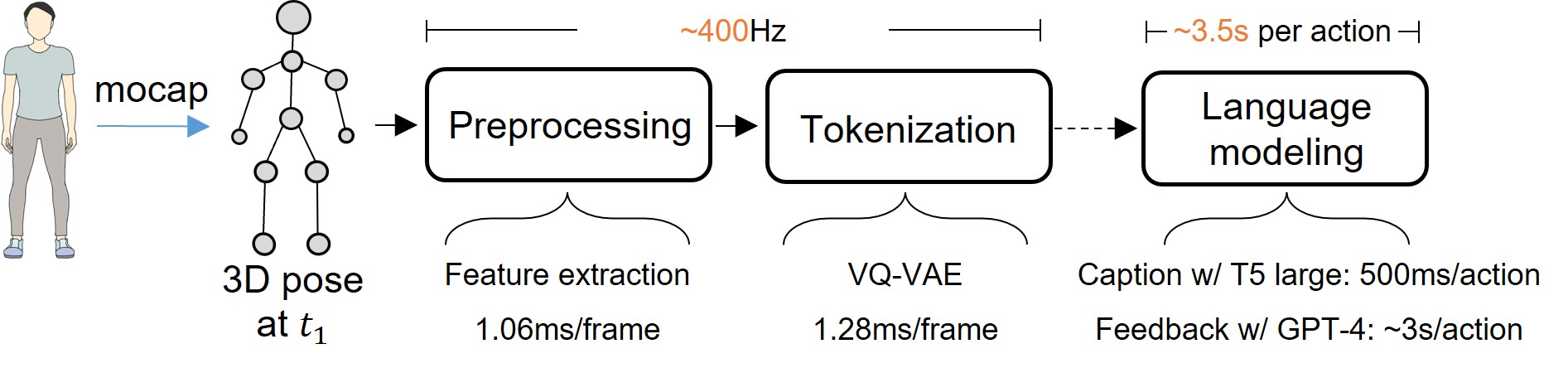}
    \caption{The inference pipeline of the proposed framework, inference time is computed based on a machine with RTX 4090.}
    \Description{}
    \label{fig:fig11}
\end{figure}

    \subsubsection{Inference Time} From the collection of 3D pose data to the generation of action description and finally the feedback, the complete inference pipeline is shown in Fig~\ref{fig:fig11}. Given our environment of RTX 4090 and a CPU with 2.30 GHz frequency, the time spends on data preprocessing and action tokenization takes 1.06ms and 1.28ms per each frame in an action data sequence, respectively. While the actual operation of the algorithm will use the multi-thread processing capacity of CPU and GPU, it takes nearly 2.34ms to process a single frame to transform action data into tokens, equivalent to processing 427 frames per second. Given the action data in an arbitrary duration, the length of action tokens range from 0 to 512 (with codebook size set to 512), while the processing time for action description stays almost the same. For T5 large, our machine needs 0.5s to provide a complete description per action instance. Meanwhile, the generation of feedback using the GPT-4 API costs approximately 3 seconds before returning responses. In short, given an action lasting 10 seconds at a streaming frequency of 60Hz, our system requires 4.91s to generate feedback for the user in a real-time setting, using a machine with RTX 4090. This duration is comparable to the time it takes a physio to make a suggestion in a clinical setting, according to the physios. However, if instant feedback for each action instance is not imperative, the longer latency associated with more commonly used computing devices may also be acceptable.

    \subsubsection{Visual Motion Capture as an Alternative to IMUs} In our experiment, in order to improve the comfort of users when wearing the IMUs, we designed a thin, stretchable, and breathable suit with wireless sensors directly attached. Although our participants generally found the suit comfortable, requiring users to have such equipment for home use is costly. This cost is not only financial but also practical, due to drifting issues that cause pose flaws over extended periods, typically beyond 10 minutes. This latter problem is impactful, which led to the exclusion of data from 15 participants in our experiment. Therefore, it is beneficial to explore pose estimation methods with alternative modalities, which could improve the realistic application of UbiPhysio.
    
    Here, we run an extra experiment with the most recent technology of acquiring 3D human poses from a video. Particularly, such a visual system may accelerate real-world deployment, eliminating the need for users to purchase and wear extra devices. We use the plug-and-play PoseFormer V2 \cite{Zhao_2023_CVPR}, applying it to videos collected at 65\degree from two randomly selected participants in the test fold. Given the common 17-join skeleton data returned by this kind of method, we simulate the joint data of head end, both hands, and feet using their nearest joint and the distance $d$ in-between, e.g., using the known coordinate of left forearm $A$, the coordinate of left hand $B$ that is missing in the visual poses is simulated as $(\hat{x},\hat{y},\hat{z})=(x,y,z) + \frac{\Vec{AB}}{|AB|}\times d$. The joint of neck 1 is acquired by taking an average of the joints of neck and head. Such that, we transform the 17-joint visual pose data into the skeleton used in this work, and put into the inference pipeline shown above. It is important to note that the model used in this experiment was pre-trained solely on data collected from the IMUs. The preprocessing conducted on the visual poses remains the same as those applied to IMU poses, including Z-normalization (with their respective distributions), skeleton normalization, and alignment towards the Z+ direction. 

\begin{figure}[t]
    \centering
    \includegraphics[width=\linewidth]{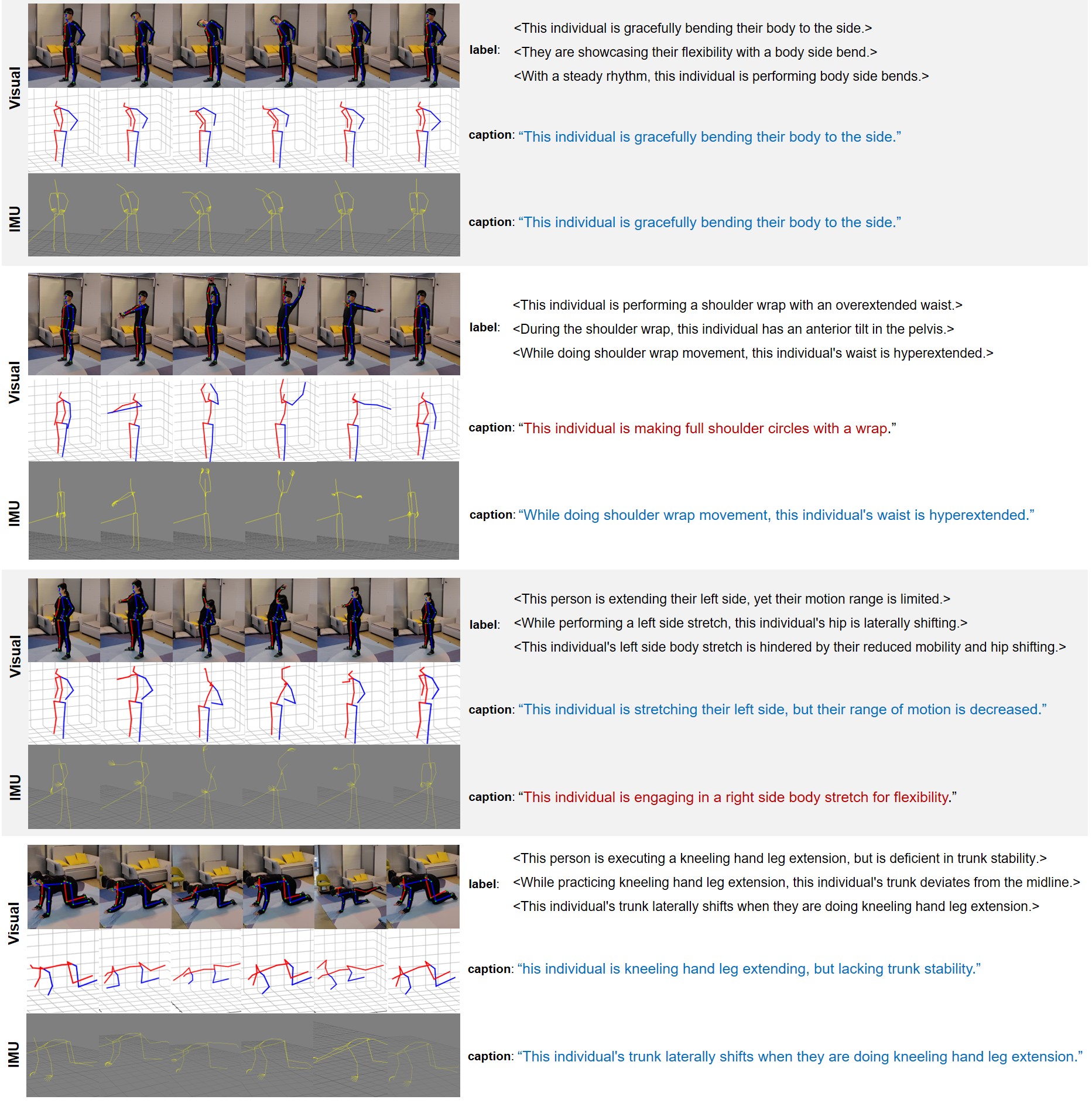}
    \caption{Visualizations of the results acquired on action data collected from a visual system and IMUs. The model's outputs are marked in blue for correct predictions and red for incorrect ones.}
    \Description{}
    \label{fig:fig12}
\end{figure}

    The results, as are shown in Fig~\ref{fig:fig12} with visualization examples, look promising and comparable to those using data from IMUs. However, for movement patterns that describe the behavior of localized body parts, such as waist hyperextension during shoulder wrap, the pose estimated with the visual system is restricted by its fidelity. In addition to the common issue of occlusions when using a single camera, this is another problem with visual systems: the shape change caused by subtle movements of the body (e.g., the waist in this case) may not get properly tracked by the visual algorithm. As for other movements, vision-captured poses provide sufficient information for our framework, leading to satisfactory results. We posit that by further refining the visual system—for instance, integrating it with signals from a few IMUs, or using high-resolution RGB-D cameras—we can enhance performance.

\section{Opportunities and Future Development}

    During our study, we found the following opportunities that may guide the future development of UbiPhysio.

\subsection{Enriching Feedback Generation via Real-time Physiological Signal Monitoring}

    In our future application scenarios involving daily functioning, fitness, and rehabilitation activities, the user's physiological signals, such as heart rate, breath rate, and heart rate variability (HRV), are essential measurements complementary to the exterior human action features. These physiological signals could serve as direct indicators of the user's internal state, such as physical load and fatigue level. Therefore, combining physiological signals with motion capture data could help the system establish more comprehensive user profiles, which, in turn, can enhance the generation of appropriate intervention feedback. As an initial exploration, We re-examined the auxiliary audio data collected from our main experiment sessions (with a wireless collar clip microphone) to figure out whether behaviors indicating the physiological state (e.g., heavy breaths) can be detected. We manually labeled 3737 audio clips from 40 users with one of the three labels - ``heavy breath, sigh, or moan'', ``talking or friction noise'', and ``background or microphone noise'' - by reviewing the corresponding video clips. The initial results suggest that the signal-to-noise ratio (SNR) of breaths vs. more salient signals like talking or background noise is not sufficient for robust recognition in our current hardware settings (e.g., breath typically has a lower amplitude and is easily overwhelmed by friction noise and background noise). To improve the recognition accuracy and robustness, one possible solution is to optimize the hardware form (e.g., using the inner microphone from an ANC earbud \cite{10.1145/3544548.3581008}). In future research, we aim to incorporate alternative sensor channels, such as PPG, RPPG, Wi-Fi, and mm wave radar, into our UbiPhysio framework to enhance the sensing capability for both motional signals and physiological signals. In this context, how these signals jointly influence the intervention and rehabilitation strategy would be an interesting research question for further investigation.

\subsection{From One-Shot Feedback to the Interactive Functioning}

    The current framework provides action description for each action instance, but it has not yet been tested in the context of continuous user interaction. For one thing, the system needs to segment the data stream into meaningful action instances, or one should verify by putting continuous data into the pipeline the language output is still useable. For another, in real-life scenarios, physios would typically respond to the performance of the user consistently, providing not only feedback, but also encouragements and other forms of verbal support. Particularly, in cases where there is a discrepancy in action descriptions, a multi-round dialogue with the user would be conducted to discern their potential physical limitations and cognitive biases. We are optimistic that it is close when an artificial intelligence agent can deliver such a comprehensive service to users. Starting from the capacity presented by our proposed UbiPhysio framework, the next step on enabling these interactive functions, e.g., to bring users in the loop, can open further opportunities. First, the interaction can provide a large amount of data, which may either be used to fine-tune the model on unseen actions and patterns or providing personalized services. Second, understanding action semantics is a new functional basis, like the role played by traditional human activity recognition (HAR), and an interactive agent built on this may facilitate impactful and interesting applications, e.g., monitoring the development of motion-affected disease \cite{ricotti2023wearable,kadirvelu2023wearable,schalkamp2023wearable}, and serving as a babysitter to prevent children from dangerous actions.

\section{Conclusion}

    In this paper, we proposed UbiPhysio, a pioneering framework designed to offer fine-grained action descriptions, in terms of the action type and movement patterns, and feedback for daily functioning, fitness, and rehabilitation activities. We have assessed the framework with extensive experiments involving 104 diverse participants who engaged in 25 types of everyday activities and exercises in a home-like setting. The quality of language output, evaluated using standard benchmarks under various tuning strategies, showed promising results. Additionally, we carried out a user study involving clinical physios and non-professional users, whose results further affirmed the framework's suitability and usability. We also explored the feasibility of deploying our framework in real-world settings using vision-captured motion data, instead of relying on the data returned by many IMUs. With these promising results, we believe that UbiPhysio can significantly enhance the effectiveness of remote fitness and rehabilitation programs, improve user engagement, and ultimately contribute to better health outcomes.


\begin{acks}
This work is supported by the Natural Science Foundation of China under Grant No. 62132010, and by Beijing Municipal Science \& Technology Commission, Administrative Commission of Zhongguancun Science Park No.Z221100006722018. We also thank Zhuo Wang, Jing Hu, and Chong Li from WCH, SCU for their active involvement during our data annotation process, and Weiwei Gao and Xinyu Xu for their brilliant visual works.
\end{acks}

\bibliographystyle{ACM-Reference-Format}
\bibliography{sample-base}


\begin{thebibliography}{72}


\ifx \showCODEN    \undefined \def \showCODEN     #1{\unskip}     \fi
\ifx \showDOI      \undefined \def \showDOI       #1{#1}\fi
\ifx \showISBNx    \undefined \def \showISBNx     #1{\unskip}     \fi
\ifx \showISBNxiii \undefined \def \showISBNxiii  #1{\unskip}     \fi
\ifx \showISSN     \undefined \def \showISSN      #1{\unskip}     \fi
\ifx \showLCCN     \undefined \def \showLCCN      #1{\unskip}     \fi
\ifx \shownote     \undefined \def \shownote      #1{#1}          \fi
\ifx \showarticletitle \undefined \def \showarticletitle #1{#1}   \fi
\ifx \showURL      \undefined \def \showURL       {\relax}        \fi
\providecommand\bibfield[2]{#2}
\providecommand\bibinfo[2]{#2}
\providecommand\natexlab[1]{#1}
\providecommand\showeprint[2][]{arXiv:#2}

\bibitem[Kin(nect)]%
        {Kinect}
 \bibinfo{year}{Kinect}\natexlab{}.
\newblock \bibinfo{title}{2023}.
\newblock
\newblock
\urldef\tempurl%
\url{www.xbox.com/en-US/kinect}
\showURL{%
\tempurl}


\bibitem[noi(itom)]%
        {noitom}
 \bibinfo{year}{Noitom}\natexlab{}.
\newblock \bibinfo{title}{2023}.
\newblock
\newblock
\urldef\tempurl%
\url{www.noitom.com/perception-neuron-series}
\showURL{%
\tempurl}


\bibitem[Aviezer et~al\mbox{.}(2012)]%
        {aviezer2012body}
\bibfield{author}{\bibinfo{person}{Hillel Aviezer}, \bibinfo{person}{Yaacov Trope}, {and} \bibinfo{person}{Alexander Todorov}.} \bibinfo{year}{2012}\natexlab{}.
\newblock \showarticletitle{Body cues, not facial expressions, discriminate between intense positive and negative emotions}.
\newblock \bibinfo{journal}{\emph{Science}} \bibinfo{volume}{338}, \bibinfo{number}{6111} (\bibinfo{year}{2012}), \bibinfo{pages}{1225--1229}.
\newblock


\bibitem[Barrett et~al\mbox{.}(2011)]%
        {barrett2011context}
\bibfield{author}{\bibinfo{person}{Lisa~Feldman Barrett}, \bibinfo{person}{Batja Mesquita}, {and} \bibinfo{person}{Maria Gendron}.} \bibinfo{year}{2011}\natexlab{}.
\newblock \showarticletitle{Context in emotion perception}.
\newblock \bibinfo{journal}{\emph{Current directions in psychological science}} \bibinfo{volume}{20}, \bibinfo{number}{5} (\bibinfo{year}{2011}), \bibinfo{pages}{286--290}.
\newblock


\bibitem[Bhattacharya et~al\mbox{.}(2022)]%
        {bhattacharya2022leveraging}
\bibfield{author}{\bibinfo{person}{Sarnab Bhattacharya}, \bibinfo{person}{Rebecca Adaimi}, {and} \bibinfo{person}{Edison Thomaz}.} \bibinfo{year}{2022}\natexlab{}.
\newblock \showarticletitle{Leveraging sound and wrist motion to detect activities of daily living with commodity smartwatches}.
\newblock \bibinfo{journal}{\emph{Proceedings of the ACM on Interactive, Mobile, Wearable and Ubiquitous Technologies}} \bibinfo{volume}{6}, \bibinfo{number}{2} (\bibinfo{year}{2022}), \bibinfo{pages}{1--28}.
\newblock


\bibitem[Biely et~al\mbox{.}(2014)]%
        {biely2014clinical}
\bibfield{author}{\bibinfo{person}{Scott~A Biely}, \bibinfo{person}{Sheri~P Silfies}, \bibinfo{person}{Susan~S Smith}, {and} \bibinfo{person}{Gregory~E Hicks}.} \bibinfo{year}{2014}\natexlab{}.
\newblock \showarticletitle{Clinical observation of standing trunk movements: What do the aberrant movement patterns tell us?}
\newblock \bibinfo{journal}{\emph{journal of orthopaedic \& sports physical therapy}} \bibinfo{volume}{44}, \bibinfo{number}{4} (\bibinfo{year}{2014}), \bibinfo{pages}{262--272}.
\newblock


\bibitem[Burke et~al\mbox{.}(2011)]%
        {burke2011self}
\bibfield{author}{\bibinfo{person}{Lora~E Burke}, \bibinfo{person}{Jing Wang}, {and} \bibinfo{person}{Mary~Ann Sevick}.} \bibinfo{year}{2011}\natexlab{}.
\newblock \showarticletitle{Self-monitoring in weight loss: a systematic review of the literature}.
\newblock \bibinfo{journal}{\emph{Journal of the American Dietetic Association}} \bibinfo{volume}{111}, \bibinfo{number}{1} (\bibinfo{year}{2011}), \bibinfo{pages}{92--102}.
\newblock


\bibitem[Chen and Xiao(2022)]%
        {chen2022harnessing}
\bibfield{author}{\bibinfo{person}{Jiangjie Chen} {and} \bibinfo{person}{Yanghua Xiao}.} \bibinfo{year}{2022}\natexlab{}.
\newblock \showarticletitle{Harnessing Knowledge and Reasoning for Human-Like Natural Language Generation: A Brief Review}.
\newblock \bibinfo{journal}{\emph{arXiv preprint arXiv:2212.03747}} (\bibinfo{year}{2022}).
\newblock


\bibitem[Chung et~al\mbox{.}(2022)]%
        {chung2022scaling}
\bibfield{author}{\bibinfo{person}{Hyung~Won Chung}, \bibinfo{person}{Le Hou}, \bibinfo{person}{Shayne Longpre}, \bibinfo{person}{Barret Zoph}, \bibinfo{person}{Yi Tay}, \bibinfo{person}{William Fedus}, \bibinfo{person}{Eric Li}, \bibinfo{person}{Xuezhi Wang}, \bibinfo{person}{Mostafa Dehghani}, \bibinfo{person}{Siddhartha Brahma}, {et~al\mbox{.}}} \bibinfo{year}{2022}\natexlab{}.
\newblock \showarticletitle{Scaling instruction-finetuned language models}.
\newblock \bibinfo{journal}{\emph{arXiv preprint arXiv:2210.11416}} (\bibinfo{year}{2022}).
\newblock


\bibitem[Clarke et~al\mbox{.}(2020)]%
        {clarke2020reactive}
\bibfield{author}{\bibinfo{person}{Christopher Clarke}, \bibinfo{person}{Doga Cavdir}, \bibinfo{person}{Patrick Chiu}, \bibinfo{person}{Laurent Denoue}, {and} \bibinfo{person}{Don Kimber}.} \bibinfo{year}{2020}\natexlab{}.
\newblock \showarticletitle{Reactive video: adaptive video playback based on user motion for supporting physical activity}. In \bibinfo{booktitle}{\emph{Proceedings of the 33rd Annual ACM Symposium on User Interface Software and Technology}}. \bibinfo{pages}{196--208}.
\newblock


\bibitem[Cook et~al\mbox{.}(2009)]%
        {cook2009ambient}
\bibfield{author}{\bibinfo{person}{Diane~J Cook}, \bibinfo{person}{Juan~C Augusto}, {and} \bibinfo{person}{Vikramaditya~R Jakkula}.} \bibinfo{year}{2009}\natexlab{}.
\newblock \showarticletitle{Ambient intelligence: Technologies, applications, and opportunities}.
\newblock \bibinfo{journal}{\emph{Pervasive and mobile computing}} \bibinfo{volume}{5}, \bibinfo{number}{4} (\bibinfo{year}{2009}), \bibinfo{pages}{277--298}.
\newblock


\bibitem[Cook et~al\mbox{.}(2012)]%
        {cook2012casas}
\bibfield{author}{\bibinfo{person}{Diane~J Cook}, \bibinfo{person}{Aaron~S Crandall}, \bibinfo{person}{Brian~L Thomas}, {and} \bibinfo{person}{Narayanan~C Krishnan}.} \bibinfo{year}{2012}\natexlab{}.
\newblock \showarticletitle{CASAS: A smart home in a box}.
\newblock \bibinfo{journal}{\emph{Computer}} \bibinfo{volume}{46}, \bibinfo{number}{7} (\bibinfo{year}{2012}), \bibinfo{pages}{62--69}.
\newblock


\bibitem[Du et~al\mbox{.}(2022)]%
        {du2022glm}
\bibfield{author}{\bibinfo{person}{Zhengxiao Du}, \bibinfo{person}{Yujie Qian}, \bibinfo{person}{Xiao Liu}, \bibinfo{person}{Ming Ding}, \bibinfo{person}{Jiezhong Qiu}, \bibinfo{person}{Zhilin Yang}, {and} \bibinfo{person}{Jie Tang}.} \bibinfo{year}{2022}\natexlab{}.
\newblock \showarticletitle{GLM: General Language Model Pretraining with Autoregressive Blank Infilling}. In \bibinfo{booktitle}{\emph{Proceedings of the 60th Annual Meeting of the Association for Computational Linguistics (Volume 1: Long Papers)}}. \bibinfo{pages}{320--335}.
\newblock


\bibitem[Garber et~al\mbox{.}(2011)]%
        {garber2011quantity}
\bibfield{author}{\bibinfo{person}{Carol~Ewing Garber}, \bibinfo{person}{Bryan Blissmer}, \bibinfo{person}{Michael~R. Deschenes}, \bibinfo{person}{Barry~A. Franklin}, \bibinfo{person}{Michael~J. Lamonte}, \bibinfo{person}{I-Min Lee}, \bibinfo{person}{David~C. Nieman}, {and} \bibinfo{person}{David~P. Swain}.} \bibinfo{year}{2011}\natexlab{}.
\newblock \showarticletitle{Quantity and Quality of Exercise for Developing and Maintaining Cardiorespiratory, Musculoskeletal, and Neuromotor Fitness in Apparently Healthy Adults: Guidance for Prescribing Exercise}.
\newblock \bibinfo{journal}{\emph{Medicine \& Science in Sports \& Exercise}} \bibinfo{volume}{43}, \bibinfo{number}{7} (\bibinfo{date}{July} \bibinfo{year}{2011}), \bibinfo{pages}{1334--1359}.
\newblock


\bibitem[Giese and Poggio(2003)]%
        {giese2003neural}
\bibfield{author}{\bibinfo{person}{Martin~A Giese} {and} \bibinfo{person}{Tomaso Poggio}.} \bibinfo{year}{2003}\natexlab{}.
\newblock \showarticletitle{Neural mechanisms for the recognition of biological movements}.
\newblock \bibinfo{journal}{\emph{Nature Reviews Neuroscience}} \bibinfo{volume}{4}, \bibinfo{number}{3} (\bibinfo{year}{2003}), \bibinfo{pages}{179--192}.
\newblock


\bibitem[Gorji et~al\mbox{.}(2022)]%
        {gorji2022pain}
\bibfield{author}{\bibinfo{person}{Sahar~Modares Gorji}, \bibinfo{person}{Hadi Mohammadi Nia~Samakosh}, \bibinfo{person}{Peter Watt}, \bibinfo{person}{Paulo Henrique~Marchetti}, {and} \bibinfo{person}{Rafael Oliveira}.} \bibinfo{year}{2022}\natexlab{}.
\newblock \showarticletitle{Pain neuroscience education and motor control exercises versus core stability exercises on pain, disability, and balance in women with chronic low back pain}.
\newblock \bibinfo{journal}{\emph{International Journal of Environmental Research and Public Health}} \bibinfo{volume}{19}, \bibinfo{number}{5} (\bibinfo{year}{2022}), \bibinfo{pages}{2694}.
\newblock


\bibitem[Guo et~al\mbox{.}(2022a)]%
        {T2M}
\bibfield{author}{\bibinfo{person}{Chuan Guo}, \bibinfo{person}{Shihao Zou}, \bibinfo{person}{Xinxin Zuo}, \bibinfo{person}{Sen Wang}, \bibinfo{person}{Wei Ji}, \bibinfo{person}{Xingyu Li}, {and} \bibinfo{person}{Li Cheng}.} \bibinfo{year}{2022}\natexlab{a}.
\newblock \showarticletitle{Generating Diverse and Natural 3D Human Motions From Text}. In \bibinfo{booktitle}{\emph{Proceedings of the IEEE/CVF Conference on Computer Vision and Pattern Recognition (CVPR)}}. \bibinfo{pages}{5152--5161}.
\newblock


\bibitem[Guo et~al\mbox{.}(2022b)]%
        {TM2T}
\bibfield{author}{\bibinfo{person}{Chuan Guo}, \bibinfo{person}{Xinxin Zuo}, \bibinfo{person}{Sen Wang}, {and} \bibinfo{person}{Li Cheng}.} \bibinfo{year}{2022}\natexlab{b}.
\newblock \showarticletitle{TM2T: Stochastic and Tokenized Modeling for the Reciprocal Generation of 3D Human Motions and Texts}. In \bibinfo{booktitle}{\emph{ECCV}}.
\newblock


\bibitem[Guo et~al\mbox{.}(2018)]%
        {guo2018device}
\bibfield{author}{\bibinfo{person}{Xiaonan Guo}, \bibinfo{person}{Jian Liu}, \bibinfo{person}{Cong Shi}, \bibinfo{person}{Hongbo Liu}, \bibinfo{person}{Yingying Chen}, {and} \bibinfo{person}{Mooi~Choo Chuah}.} \bibinfo{year}{2018}\natexlab{}.
\newblock \showarticletitle{Device-free personalized fitness assistant using WiFi}.
\newblock \bibinfo{journal}{\emph{Proceedings of the ACM on Interactive, Mobile, Wearable and Ubiquitous Technologies}} \bibinfo{volume}{2}, \bibinfo{number}{4} (\bibinfo{year}{2018}), \bibinfo{pages}{1--23}.
\newblock


\bibitem[Guti{\'e}rrez-Espinoza et~al\mbox{.}(2022)]%
        {gutierrez2022effectiveness}
\bibfield{author}{\bibinfo{person}{H{\'e}ctor Guti{\'e}rrez-Espinoza}, \bibinfo{person}{Felipe Araya-Quintanilla}, \bibinfo{person}{Cristian Olgu{\'\i}n-Huerta}, \bibinfo{person}{Iv{\'a}n Vald{\'e}s-Orrego}, {and} \bibinfo{person}{Oscar Sep{\'u}lveda-Osses}.} \bibinfo{year}{2022}\natexlab{}.
\newblock \showarticletitle{Effectiveness of supervised physiotherapy versus home exercise in subjects with rotator cuff disorders treated surgically: A systematic review and meta-analysis}.
\newblock \bibinfo{journal}{\emph{Physiotherapy Research International}} \bibinfo{volume}{27}, \bibinfo{number}{2} (\bibinfo{year}{2022}).
\newblock


\bibitem[Guu et~al\mbox{.}(2020)]%
        {guu2020retrieval}
\bibfield{author}{\bibinfo{person}{Kelvin Guu}, \bibinfo{person}{Kenton Lee}, \bibinfo{person}{Zora Tung}, \bibinfo{person}{Panupong Pasupat}, {and} \bibinfo{person}{Mingwei Chang}.} \bibinfo{year}{2020}\natexlab{}.
\newblock \showarticletitle{Retrieval augmented language model pre-training}. In \bibinfo{booktitle}{\emph{International conference on machine learning}}. PMLR, \bibinfo{pages}{3929--3938}.
\newblock


\bibitem[Halliday et~al\mbox{.}(2019)]%
        {halliday2019randomized}
\bibfield{author}{\bibinfo{person}{Mark~H Halliday}, \bibinfo{person}{Evangelos Pappas}, \bibinfo{person}{Mark~J Hancock}, \bibinfo{person}{Helen~A Clare}, \bibinfo{person}{Rafael~Z Pinto}, \bibinfo{person}{Gavin Robertson}, {and} \bibinfo{person}{Paulo~H Ferreira}.} \bibinfo{year}{2019}\natexlab{}.
\newblock \showarticletitle{A randomized clinical trial comparing the McKenzie method and motor control exercises in people with chronic low back pain and a directional preference: 1-year follow-up}.
\newblock \bibinfo{journal}{\emph{Physiotherapy}} \bibinfo{volume}{105}, \bibinfo{number}{4} (\bibinfo{year}{2019}), \bibinfo{pages}{442--445}.
\newblock


\bibitem[Halloran et~al\mbox{.}(2019)]%
        {halloran2019remote}
\bibfield{author}{\bibinfo{person}{Shane Halloran}, \bibinfo{person}{Lin Tang}, \bibinfo{person}{Yu Guan}, \bibinfo{person}{Jian~Qing Shi}, {and} \bibinfo{person}{Janet Eyre}.} \bibinfo{year}{2019}\natexlab{}.
\newblock \showarticletitle{Remote monitoring of stroke patients' rehabilitation using wearable accelerometers}. In \bibinfo{booktitle}{\emph{Proceedings of the 2019 ACM International Symposium on Wearable Computers}}. \bibinfo{pages}{72--77}.
\newblock


\bibitem[Haresamudram et~al\mbox{.}(2023)]%
        {haresamudram2023towards}
\bibfield{author}{\bibinfo{person}{Harish Haresamudram}, \bibinfo{person}{Irfan Essa}, {and} \bibinfo{person}{Thomas Ploetz}.} \bibinfo{year}{2023}\natexlab{}.
\newblock \showarticletitle{Towards Learning Discrete Representations via Self-Supervision for Wearables-Based Human Activity Recognition}.
\newblock \bibinfo{journal}{\emph{arXiv preprint arXiv:2306.01108}} (\bibinfo{year}{2023}).
\newblock


\bibitem[Hiremath et~al\mbox{.}(2022)]%
        {hiremath2022bootstrapping}
\bibfield{author}{\bibinfo{person}{Shruthi~K Hiremath}, \bibinfo{person}{Yasutaka Nishimura}, \bibinfo{person}{Sonia Chernova}, {and} \bibinfo{person}{Thomas Pl{\"o}tz}.} \bibinfo{year}{2022}\natexlab{}.
\newblock \showarticletitle{Bootstrapping Human Activity Recognition Systems for Smart Homes from Scratch}.
\newblock \bibinfo{journal}{\emph{Proceedings of the ACM on Interactive, Mobile, Wearable and Ubiquitous Technologies}} \bibinfo{volume}{6}, \bibinfo{number}{3} (\bibinfo{year}{2022}), \bibinfo{pages}{1--27}.
\newblock


\bibitem[Hoang et~al\mbox{.}(2016)]%
        {hoang2016onebody}
\bibfield{author}{\bibinfo{person}{Thuong~N Hoang}, \bibinfo{person}{Martin Reinoso}, \bibinfo{person}{Frank Vetere}, {and} \bibinfo{person}{Egemen Tanin}.} \bibinfo{year}{2016}\natexlab{}.
\newblock \showarticletitle{Onebody: remote posture guidance system using first person view in virtual environment}. In \bibinfo{booktitle}{\emph{Proceedings of the 9th Nordic Conference on Human-Computer Interaction}}. \bibinfo{pages}{1--10}.
\newblock


\bibitem[Holden et~al\mbox{.}(2020)]%
        {feature2}
\bibfield{author}{\bibinfo{person}{Daniel Holden}, \bibinfo{person}{Oussama Kanoun}, \bibinfo{person}{Maksym Perepichka}, {and} \bibinfo{person}{Tiberiu Popa}.} \bibinfo{year}{2020}\natexlab{}.
\newblock \showarticletitle{Learned Motion Matching}.
\newblock \bibinfo{journal}{\emph{ACM Trans. Graph.}} \bibinfo{volume}{39}, \bibinfo{number}{4}, Article \bibinfo{articleno}{53} (\bibinfo{date}{aug} \bibinfo{year}{2020}), \bibinfo{numpages}{13}~pages.
\newblock
\showISSN{0730-0301}


\bibitem[Holden et~al\mbox{.}(2017)]%
        {feature1}
\bibfield{author}{\bibinfo{person}{Daniel Holden}, \bibinfo{person}{Taku Komura}, {and} \bibinfo{person}{Jun Saito}.} \bibinfo{year}{2017}\natexlab{}.
\newblock \showarticletitle{Phase-Functioned Neural Networks for Character Control}.
\newblock \bibinfo{journal}{\emph{ACM Trans. Graph.}} \bibinfo{volume}{36}, \bibinfo{number}{4}, Article \bibinfo{articleno}{42} (\bibinfo{date}{jul} \bibinfo{year}{2017}), \bibinfo{numpages}{13}~pages.
\newblock
\showISSN{0730-0301}


\bibitem[Hu et~al\mbox{.}(2021)]%
        {hu2021lora}
\bibfield{author}{\bibinfo{person}{Edward~J Hu}, \bibinfo{person}{Yelong Shen}, \bibinfo{person}{Phillip Wallis}, \bibinfo{person}{Zeyuan Allen-Zhu}, \bibinfo{person}{Yuanzhi Li}, \bibinfo{person}{Shean Wang}, \bibinfo{person}{Lu Wang}, {and} \bibinfo{person}{Weizhu Chen}.} \bibinfo{year}{2021}\natexlab{}.
\newblock \showarticletitle{Lora: Low-rank adaptation of large language models}.
\newblock \bibinfo{journal}{\emph{arXiv preprint arXiv:2106.09685}} (\bibinfo{year}{2021}).
\newblock


\bibitem[Jiang et~al\mbox{.}(2023)]%
        {jiang2023motiongpt}
\bibfield{author}{\bibinfo{person}{Biao Jiang}, \bibinfo{person}{Xin Chen}, \bibinfo{person}{Wen Liu}, \bibinfo{person}{Jingyi Yu}, \bibinfo{person}{Gang Yu}, {and} \bibinfo{person}{Tao Chen}.} \bibinfo{year}{2023}\natexlab{}.
\newblock \showarticletitle{MotionGPT: Human Motion as a Foreign Language}.
\newblock \bibinfo{journal}{\emph{arXiv preprint arXiv:2306.14795}} (\bibinfo{year}{2023}).
\newblock


\bibitem[Jo et~al\mbox{.}(2023)]%
        {jo2023flowar}
\bibfield{author}{\bibinfo{person}{Hye-Young Jo}, \bibinfo{person}{Laurenz Seidel}, \bibinfo{person}{Michel Pahud}, \bibinfo{person}{Mike Sinclair}, {and} \bibinfo{person}{Andrea Bianchi}.} \bibinfo{year}{2023}\natexlab{}.
\newblock \showarticletitle{FlowAR: How Different Augmented Reality Visualizations of Online Fitness Videos Support Flow for At-Home Yoga Exercises}. In \bibinfo{booktitle}{\emph{Proceedings of the 2023 CHI Conference on Human Factors in Computing Systems}}. \bibinfo{pages}{1--17}.
\newblock


\bibitem[Kadirvelu et~al\mbox{.}(2023)]%
        {kadirvelu2023wearable}
\bibfield{author}{\bibinfo{person}{Balasundaram Kadirvelu}, \bibinfo{person}{Constantinos Gavriel}, \bibinfo{person}{Sathiji Nageshwaran}, \bibinfo{person}{Jackson Ping~Kei Chan}, \bibinfo{person}{Suran Nethisinghe}, \bibinfo{person}{Stavros Athanasopoulos}, \bibinfo{person}{Valeria Ricotti}, \bibinfo{person}{Thomas Voit}, \bibinfo{person}{Paola Giunti}, \bibinfo{person}{Richard Festenstein}, {et~al\mbox{.}}} \bibinfo{year}{2023}\natexlab{}.
\newblock \showarticletitle{A wearable motion capture suit and machine learning predict disease progression in Friedreich’s ataxia}.
\newblock \bibinfo{journal}{\emph{Nature Medicine}} \bibinfo{volume}{29}, \bibinfo{number}{1} (\bibinfo{year}{2023}), \bibinfo{pages}{86--94}.
\newblock


\bibitem[Kim and Yim(2020)]%
        {kim2020core}
\bibfield{author}{\bibinfo{person}{Beomryong Kim} {and} \bibinfo{person}{Jongeun Yim}.} \bibinfo{year}{2020}\natexlab{}.
\newblock \showarticletitle{Core stability and hip exercises improve physical function and activity in patients with non-specific low back pain: a randomized controlled trial}.
\newblock \bibinfo{journal}{\emph{The Tohoku journal of experimental medicine}} \bibinfo{volume}{251}, \bibinfo{number}{3} (\bibinfo{year}{2020}), \bibinfo{pages}{193--206}.
\newblock


\bibitem[Kim et~al\mbox{.}(2022)]%
        {kim2022learning}
\bibfield{author}{\bibinfo{person}{Jihoon Kim}, \bibinfo{person}{Youngjae Yu}, \bibinfo{person}{Seungyoun Shin}, \bibinfo{person}{Taehyun Byun}, {and} \bibinfo{person}{Sungjoon Choi}.} \bibinfo{year}{2022}\natexlab{}.
\newblock \bibinfo{title}{Learning Joint Representation of Human Motion and Language}.
\newblock
\newblock
\showeprint[arxiv]{2210.15187}


\bibitem[Law et~al\mbox{.}(2014)]%
        {law2014effects}
\bibfield{author}{\bibinfo{person}{Lawla~LF Law}, \bibinfo{person}{Fiona Barnett}, \bibinfo{person}{Matthew~K Yau}, {and} \bibinfo{person}{Marion~A Gray}.} \bibinfo{year}{2014}\natexlab{}.
\newblock \showarticletitle{Effects of functional tasks exercise on older adults with cognitive impairment at risk of Alzheimer's disease: a randomised controlled trial}.
\newblock \bibinfo{journal}{\emph{Age and ageing}} \bibinfo{volume}{43}, \bibinfo{number}{6} (\bibinfo{year}{2014}), \bibinfo{pages}{813--820}.
\newblock


\bibitem[Lee et~al\mbox{.}(2021)]%
        {lee2021human}
\bibfield{author}{\bibinfo{person}{Min~Hun Lee}, \bibinfo{person}{Daniel~P Siewiorek}, \bibinfo{person}{Asim Smailagic}, \bibinfo{person}{Alexandre Bernardino}, {and} \bibinfo{person}{Sergi Berm{\'u}dez~i Badia}.} \bibinfo{year}{2021}\natexlab{}.
\newblock \showarticletitle{A human-ai collaborative approach for clinical decision making on rehabilitation assessment}. In \bibinfo{booktitle}{\emph{Proceedings of the 2021 CHI conference on human factors in computing systems}}. \bibinfo{pages}{1--14}.
\newblock


\bibitem[Li et~al\mbox{.}(2023)]%
        {10.1145/3544548.3581008}
\bibfield{author}{\bibinfo{person}{Zisu Li}, \bibinfo{person}{Chen Liang}, \bibinfo{person}{Yuntao Wang}, \bibinfo{person}{Yue Qin}, \bibinfo{person}{Chun Yu}, \bibinfo{person}{Yukang Yan}, \bibinfo{person}{Mingming Fan}, {and} \bibinfo{person}{Yuanchun Shi}.} \bibinfo{year}{2023}\natexlab{}.
\newblock \showarticletitle{Enabling Voice-Accompanying Hand-to-Face Gesture Recognition with Cross-Device Sensing}. In \bibinfo{booktitle}{\emph{Proceedings of the 2023 CHI Conference on Human Factors in Computing Systems}} (Hamburg, Germany) \emph{(\bibinfo{series}{CHI '23})}. \bibinfo{publisher}{Association for Computing Machinery}, \bibinfo{address}{New York, NY, USA}, Article \bibinfo{articleno}{313}, \bibinfo{numpages}{17}~pages.
\newblock
\showISBNx{9781450394215}


\bibitem[Liao et~al\mbox{.}(2020)]%
        {liao2020review}
\bibfield{author}{\bibinfo{person}{Yalin Liao}, \bibinfo{person}{Aleksandar Vakanski}, \bibinfo{person}{Min Xian}, \bibinfo{person}{David Paul}, {and} \bibinfo{person}{Russell Baker}.} \bibinfo{year}{2020}\natexlab{}.
\newblock \showarticletitle{A review of computational approaches for evaluation of rehabilitation exercises}.
\newblock \bibinfo{journal}{\emph{Computers in biology and medicine}}  \bibinfo{volume}{119} (\bibinfo{year}{2020}), \bibinfo{pages}{103687}.
\newblock


\bibitem[Lin(2004)]%
        {lin2004rouge}
\bibfield{author}{\bibinfo{person}{Chin-Yew Lin}.} \bibinfo{year}{2004}\natexlab{}.
\newblock \showarticletitle{Rouge: A package for automatic evaluation of summaries}. In \bibinfo{booktitle}{\emph{Text summarization branches out}}. \bibinfo{pages}{74--81}.
\newblock


\bibitem[Liu et~al\mbox{.}(2022)]%
        {liu2022ptuning}
\bibfield{author}{\bibinfo{person}{Xiao Liu}, \bibinfo{person}{Kaixuan Ji}, \bibinfo{person}{Yicheng Fu}, \bibinfo{person}{Weng Tam}, \bibinfo{person}{Zhengxiao Du}, \bibinfo{person}{Zhilin Yang}, {and} \bibinfo{person}{Jie Tang}.} \bibinfo{year}{2022}\natexlab{}.
\newblock \showarticletitle{P-tuning: Prompt tuning can be comparable to fine-tuning across scales and tasks}. In \bibinfo{booktitle}{\emph{Proceedings of the 60th Annual Meeting of the Association for Computational Linguistics (Volume 2: Short Papers)}}. \bibinfo{pages}{61--68}.
\newblock


\bibitem[Loper et~al\mbox{.}(2023)]%
        {10.1145/3596711.3596800}
\bibfield{author}{\bibinfo{person}{Matthew Loper}, \bibinfo{person}{Naureen Mahmood}, \bibinfo{person}{Javier Romero}, \bibinfo{person}{Gerard Pons-Moll}, {and} \bibinfo{person}{Michael~J. Black}.} \bibinfo{year}{2023}\natexlab{}.
\newblock \bibinfo{booktitle}{\emph{SMPL: A Skinned Multi-Person Linear Model} (\bibinfo{edition}{1} ed.)}.
\newblock \bibinfo{publisher}{Association for Computing Machinery}, \bibinfo{address}{New York, NY, USA}.
\newblock
\showISBNx{9798400708978}


\bibitem[Loshchilov and Hutter(2017)]%
        {loshchilov2017decoupled}
\bibfield{author}{\bibinfo{person}{Ilya Loshchilov} {and} \bibinfo{person}{Frank Hutter}.} \bibinfo{year}{2017}\natexlab{}.
\newblock \showarticletitle{Decoupled weight decay regularization}.
\newblock \bibinfo{journal}{\emph{arXiv preprint arXiv:1711.05101}} (\bibinfo{year}{2017}).
\newblock


\bibitem[Major et~al\mbox{.}(2021)]%
        {major2021feasibility}
\bibfield{author}{\bibinfo{person}{Mel~E Major}, \bibinfo{person}{Daniela Dettling-Ihnenfeldt}, \bibinfo{person}{Stephan~PJ Ramaekers}, \bibinfo{person}{Raoul~HH Engelbert}, {and} \bibinfo{person}{Marike van~der Schaaf}.} \bibinfo{year}{2021}\natexlab{}.
\newblock \showarticletitle{Feasibility of a home-based interdisciplinary rehabilitation program for patients with Post-Intensive Care Syndrome: the REACH study}.
\newblock \bibinfo{journal}{\emph{Critical Care}} \bibinfo{volume}{25}, \bibinfo{number}{1} (\bibinfo{year}{2021}), \bibinfo{pages}{1--15}.
\newblock


\bibitem[McLean et~al\mbox{.}(2010)]%
        {mclean2010interventions}
\bibfield{author}{\bibinfo{person}{Sionnadh~Mairi McLean}, \bibinfo{person}{Maria Burton}, \bibinfo{person}{Lesley Bradley}, {and} \bibinfo{person}{Chris Littlewood}.} \bibinfo{year}{2010}\natexlab{}.
\newblock \showarticletitle{Interventions for enhancing adherence with physiotherapy: a systematic review}.
\newblock \bibinfo{journal}{\emph{Manual therapy}} \bibinfo{volume}{15}, \bibinfo{number}{6} (\bibinfo{year}{2010}), \bibinfo{pages}{514--521}.
\newblock


\bibitem[Moschetti et~al\mbox{.}(2017)]%
        {moschetti2017daily}
\bibfield{author}{\bibinfo{person}{Alessandra Moschetti}, \bibinfo{person}{Laura Fiorini}, \bibinfo{person}{Dario Esposito}, \bibinfo{person}{Paolo Dario}, {and} \bibinfo{person}{Filippo Cavallo}.} \bibinfo{year}{2017}\natexlab{}.
\newblock \showarticletitle{Daily activity recognition with inertial ring and bracelet: An unsupervised approach}. In \bibinfo{booktitle}{\emph{2017 IEEE International Conference on Robotics and Automation (ICRA)}}. IEEE, \bibinfo{pages}{3250--3255}.
\newblock


\bibitem[Papineni et~al\mbox{.}(2002)]%
        {papineni2002bleu}
\bibfield{author}{\bibinfo{person}{Kishore Papineni}, \bibinfo{person}{Salim Roukos}, \bibinfo{person}{Todd Ward}, {and} \bibinfo{person}{Wei-Jing Zhu}.} \bibinfo{year}{2002}\natexlab{}.
\newblock \showarticletitle{Bleu: a method for automatic evaluation of machine translation}. In \bibinfo{booktitle}{\emph{Proceedings of the 40th annual meeting of the Association for Computational Linguistics}}. \bibinfo{pages}{311--318}.
\newblock


\bibitem[Piwek et~al\mbox{.}(2016)]%
        {piwek2016rise}
\bibfield{author}{\bibinfo{person}{Lukasz Piwek}, \bibinfo{person}{David~A Ellis}, \bibinfo{person}{Sally Andrews}, {and} \bibinfo{person}{Adam Joinson}.} \bibinfo{year}{2016}\natexlab{}.
\newblock \showarticletitle{The rise of consumer health wearables: promises and barriers}.
\newblock \bibinfo{journal}{\emph{PLoS medicine}} \bibinfo{volume}{13}, \bibinfo{number}{2} (\bibinfo{year}{2016}), \bibinfo{pages}{e1001953}.
\newblock


\bibitem[Plappert et~al\mbox{.}(2018)]%
        {mtmap1}
\bibfield{author}{\bibinfo{person}{Matthias Plappert}, \bibinfo{person}{Christian Mandery}, {and} \bibinfo{person}{Tamim Asfour}.} \bibinfo{year}{2018}\natexlab{}.
\newblock \showarticletitle{Learning a bidirectional mapping between human whole-body motion and natural language using deep recurrent neural networks}.
\newblock \bibinfo{journal}{\emph{Robotics and Autonomous Systems}} (\bibinfo{date}{Nov} \bibinfo{year}{2018}), \bibinfo{pages}{13–26}.
\newblock


\bibitem[Pl{\"o}tz et~al\mbox{.}(2011)]%
        {plotz2011activity}
\bibfield{author}{\bibinfo{person}{Thomas Pl{\"o}tz}, \bibinfo{person}{Paula Moynihan}, \bibinfo{person}{Cuong Pham}, {and} \bibinfo{person}{Patrick Olivier}.} \bibinfo{year}{2011}\natexlab{}.
\newblock \showarticletitle{Activity recognition and healthier food preparation}.
\newblock \bibinfo{journal}{\emph{Activity Recognition in Pervasive Intelligent Environments}} (\bibinfo{year}{2011}), \bibinfo{pages}{313--329}.
\newblock


\bibitem[Raffel et~al\mbox{.}(2020)]%
        {2020t5}
\bibfield{author}{\bibinfo{person}{Colin Raffel}, \bibinfo{person}{Noam Shazeer}, \bibinfo{person}{Adam Roberts}, \bibinfo{person}{Katherine Lee}, \bibinfo{person}{Sharan Narang}, \bibinfo{person}{Michael Matena}, \bibinfo{person}{Yanqi Zhou}, \bibinfo{person}{Wei Li}, {and} \bibinfo{person}{Peter~J. Liu}.} \bibinfo{year}{2020}\natexlab{}.
\newblock \showarticletitle{Exploring the Limits of Transfer Learning with a Unified Text-to-Text Transformer}.
\newblock \bibinfo{journal}{\emph{Journal of Machine Learning Research}} \bibinfo{volume}{21}, \bibinfo{number}{140} (\bibinfo{year}{2020}), \bibinfo{pages}{1--67}.
\newblock


\bibitem[Ram et~al\mbox{.}(2023)]%
        {ram2023context}
\bibfield{author}{\bibinfo{person}{Ori Ram}, \bibinfo{person}{Yoav Levine}, \bibinfo{person}{Itay Dalmedigos}, \bibinfo{person}{Dor Muhlgay}, \bibinfo{person}{Amnon Shashua}, \bibinfo{person}{Kevin Leyton-Brown}, {and} \bibinfo{person}{Yoav Shoham}.} \bibinfo{year}{2023}\natexlab{}.
\newblock \showarticletitle{In-context retrieval-augmented language models}.
\newblock \bibinfo{journal}{\emph{arXiv preprint arXiv:2302.00083}} (\bibinfo{year}{2023}).
\newblock


\bibitem[Razavi et~al\mbox{.}(2019)]%
        {razavi2019generating}
\bibfield{author}{\bibinfo{person}{Ali Razavi}, \bibinfo{person}{Aaron Van~den Oord}, {and} \bibinfo{person}{Oriol Vinyals}.} \bibinfo{year}{2019}\natexlab{}.
\newblock \showarticletitle{Generating diverse high-fidelity images with vq-vae-2}.
\newblock \bibinfo{journal}{\emph{Advances in neural information processing systems}}  \bibinfo{volume}{32} (\bibinfo{year}{2019}).
\newblock


\bibitem[Ricotti et~al\mbox{.}(2023)]%
        {ricotti2023wearable}
\bibfield{author}{\bibinfo{person}{Valeria Ricotti}, \bibinfo{person}{Balasundaram Kadirvelu}, \bibinfo{person}{Victoria Selby}, \bibinfo{person}{Richard Festenstein}, \bibinfo{person}{Eugenio Mercuri}, \bibinfo{person}{Thomas Voit}, {and} \bibinfo{person}{A~Aldo Faisal}.} \bibinfo{year}{2023}\natexlab{}.
\newblock \showarticletitle{Wearable full-body motion tracking of activities of daily living predicts disease trajectory in Duchenne muscular dystrophy}.
\newblock \bibinfo{journal}{\emph{Nature medicine}} \bibinfo{volume}{29}, \bibinfo{number}{1} (\bibinfo{year}{2023}), \bibinfo{pages}{95--103}.
\newblock


\bibitem[Schalkamp et~al\mbox{.}(2023)]%
        {schalkamp2023wearable}
\bibfield{author}{\bibinfo{person}{Ann-Kathrin Schalkamp}, \bibinfo{person}{Kathryn~J Peall}, \bibinfo{person}{Neil~A Harrison}, {and} \bibinfo{person}{Cynthia Sandor}.} \bibinfo{year}{2023}\natexlab{}.
\newblock \showarticletitle{Wearable movement-tracking data identify Parkinson’s disease years before clinical diagnosis}.
\newblock \bibinfo{journal}{\emph{Nature Medicine}} (\bibinfo{year}{2023}), \bibinfo{pages}{1--9}.
\newblock


\bibitem[Su et~al\mbox{.}(2014)]%
        {su2014kinect}
\bibfield{author}{\bibinfo{person}{Chuan-Jun Su}, \bibinfo{person}{Chang-Yu Chiang}, {and} \bibinfo{person}{Jing-Yan Huang}.} \bibinfo{year}{2014}\natexlab{}.
\newblock \showarticletitle{Kinect-enabled home-based rehabilitation system using Dynamic Time Warping and fuzzy logic}.
\newblock \bibinfo{journal}{\emph{Applied Soft Computing}}  \bibinfo{volume}{22} (\bibinfo{year}{2014}), \bibinfo{pages}{652--666}.
\newblock


\bibitem[Touvron et~al\mbox{.}(2023)]%
        {touvron2023llama}
\bibfield{author}{\bibinfo{person}{Hugo Touvron}, \bibinfo{person}{Louis Martin}, \bibinfo{person}{Kevin Stone}, \bibinfo{person}{Peter Albert}, \bibinfo{person}{Amjad Almahairi}, \bibinfo{person}{Yasmine Babaei}, \bibinfo{person}{Nikolay Bashlykov}, \bibinfo{person}{Soumya Batra}, \bibinfo{person}{Prajjwal Bhargava}, \bibinfo{person}{Shruti Bhosale}, {et~al\mbox{.}}} \bibinfo{year}{2023}\natexlab{}.
\newblock \showarticletitle{Llama 2: Open foundation and fine-tuned chat models}.
\newblock \bibinfo{journal}{\emph{arXiv preprint arXiv:2307.09288}} (\bibinfo{year}{2023}).
\newblock


\bibitem[Van Den~Oord et~al\mbox{.}(2017)]%
        {VQVAE}
\bibfield{author}{\bibinfo{person}{Aaron Van Den~Oord}, \bibinfo{person}{Oriol Vinyals}, {et~al\mbox{.}}} \bibinfo{year}{2017}\natexlab{}.
\newblock \showarticletitle{Neural discrete representation learning}.
\newblock \bibinfo{journal}{\emph{Advances in neural information processing systems}}  \bibinfo{volume}{30} (\bibinfo{year}{2017}).
\newblock


\bibitem[Vedantam et~al\mbox{.}(2015)]%
        {vedantam2015cider}
\bibfield{author}{\bibinfo{person}{Ramakrishna Vedantam}, \bibinfo{person}{C Lawrence~Zitnick}, {and} \bibinfo{person}{Devi Parikh}.} \bibinfo{year}{2015}\natexlab{}.
\newblock \showarticletitle{Cider: Consensus-based image description evaluation}. In \bibinfo{booktitle}{\emph{Proceedings of the IEEE conference on computer vision and pattern recognition}}. \bibinfo{pages}{4566--4575}.
\newblock


\bibitem[Wang et~al\mbox{.}(2021a)]%
        {wang2021leveraging}
\bibfield{author}{\bibinfo{person}{Chongyang Wang}, \bibinfo{person}{Yuan Gao}, \bibinfo{person}{Akhil Mathur}, \bibinfo{person}{Amanda~C De~C.~Williams}, \bibinfo{person}{Nicholas~D Lane}, {and} \bibinfo{person}{Nadia Bianchi-Berthouze}.} \bibinfo{year}{2021}\natexlab{a}.
\newblock \showarticletitle{Leveraging activity recognition to enable protective behavior detection in continuous data}.
\newblock \bibinfo{journal}{\emph{Proceedings of the ACM on Interactive, Mobile, Wearable and Ubiquitous Technologies}} \bibinfo{volume}{5}, \bibinfo{number}{2} (\bibinfo{year}{2021}), \bibinfo{pages}{1--27}.
\newblock


\bibitem[Wang et~al\mbox{.}(2019)]%
        {wang2019recurrent}
\bibfield{author}{\bibinfo{person}{Chongyang Wang}, \bibinfo{person}{Temitayo~A Olugbade}, \bibinfo{person}{Akhil Mathur}, \bibinfo{person}{Amanda~C De~C.~Williams}, \bibinfo{person}{Nicholas~D Lane}, {and} \bibinfo{person}{Nadia Bianchi-Berthouze}.} \bibinfo{year}{2019}\natexlab{}.
\newblock \showarticletitle{Recurrent network based automatic detection of chronic pain protective behavior using mocap and semg data}. In \bibinfo{booktitle}{\emph{Proceedings of the 2019 ACM International Symposium on Wearable Computers}}. \bibinfo{pages}{225--230}.
\newblock


\bibitem[Wang et~al\mbox{.}(2021b)]%
        {wang2021chronic}
\bibfield{author}{\bibinfo{person}{Chongyang Wang}, \bibinfo{person}{Temitayo~A Olugbade}, \bibinfo{person}{Akhil Mathur}, \bibinfo{person}{Amanda C DE~C Williams}, \bibinfo{person}{Nicholas~D Lane}, {and} \bibinfo{person}{Nadia Bianchi-Berthouze}.} \bibinfo{year}{2021}\natexlab{b}.
\newblock \showarticletitle{Chronic pain protective behavior detection with deep learning}.
\newblock \bibinfo{journal}{\emph{ACM Transactions on Computing for Healthcare}} \bibinfo{volume}{2}, \bibinfo{number}{3} (\bibinfo{year}{2021}), \bibinfo{pages}{1--24}.
\newblock


\bibitem[Wang and Ma(2023)]%
        {wang2023physiq}
\bibfield{author}{\bibinfo{person}{Hanchen~David Wang} {and} \bibinfo{person}{Meiyi Ma}.} \bibinfo{year}{2023}\natexlab{}.
\newblock \showarticletitle{PhysiQ: Off-site Quality Assessment of Exercise in Physical Therapy}.
\newblock \bibinfo{journal}{\emph{Proceedings of the ACM on Interactive, Mobile, Wearable and Ubiquitous Technologies}} \bibinfo{volume}{6}, \bibinfo{number}{4} (\bibinfo{year}{2023}), \bibinfo{pages}{1--25}.
\newblock


\bibitem[Wang et~al\mbox{.}(2003)]%
        {wang2003recent}
\bibfield{author}{\bibinfo{person}{Liang Wang}, \bibinfo{person}{Weiming Hu}, {and} \bibinfo{person}{Tieniu Tan}.} \bibinfo{year}{2003}\natexlab{}.
\newblock \showarticletitle{Recent developments in human motion analysis}.
\newblock \bibinfo{journal}{\emph{Pattern recognition}} \bibinfo{volume}{36}, \bibinfo{number}{3} (\bibinfo{year}{2003}), \bibinfo{pages}{585--601}.
\newblock


\bibitem[Wang et~al\mbox{.}(2022)]%
        {NEURIPS2022_6030db51}
\bibfield{author}{\bibinfo{person}{Zan Wang}, \bibinfo{person}{Yixin Chen}, \bibinfo{person}{Tengyu Liu}, \bibinfo{person}{Yixin Zhu}, \bibinfo{person}{Wei Liang}, {and} \bibinfo{person}{Siyuan Huang}.} \bibinfo{year}{2022}\natexlab{}.
\newblock \showarticletitle{HUMANISE: Language-conditioned Human Motion Generation in 3D Scenes}. In \bibinfo{booktitle}{\emph{Advances in Neural Information Processing Systems}}, \bibfield{editor}{\bibinfo{person}{S.~Koyejo}, \bibinfo{person}{S.~Mohamed}, \bibinfo{person}{A.~Agarwal}, \bibinfo{person}{D.~Belgrave}, \bibinfo{person}{K.~Cho}, {and} \bibinfo{person}{A.~Oh}} (Eds.), Vol.~\bibinfo{volume}{35}. \bibinfo{publisher}{Curran Associates, Inc.}, \bibinfo{pages}{14959--14971}.
\newblock


\bibitem[Wei et~al\mbox{.}(2019)]%
        {wei2019towards}
\bibfield{author}{\bibinfo{person}{Wenchuan Wei}, \bibinfo{person}{Carter McElroy}, {and} \bibinfo{person}{Sujit Dey}.} \bibinfo{year}{2019}\natexlab{}.
\newblock \showarticletitle{Towards on-demand virtual physical therapist: Machine learning-based patient action understanding, assessment and task recommendation}.
\newblock \bibinfo{journal}{\emph{IEEE Transactions on Neural Systems and Rehabilitation Engineering}} \bibinfo{volume}{27}, \bibinfo{number}{9} (\bibinfo{year}{2019}), \bibinfo{pages}{1824--1835}.
\newblock


\bibitem[Xie et~al\mbox{.}(2021)]%
        {xie2021hearfit+}
\bibfield{author}{\bibinfo{person}{Yadong Xie}, \bibinfo{person}{Fan Li}, \bibinfo{person}{Yue Wu}, {and} \bibinfo{person}{Yu Wang}.} \bibinfo{year}{2021}\natexlab{}.
\newblock \showarticletitle{HearFit+: Personalized fitness monitoring via audio signals on smart speakers}.
\newblock \bibinfo{journal}{\emph{IEEE Transactions on Mobile Computing}} (\bibinfo{year}{2021}).
\newblock


\bibitem[Yamada et~al\mbox{.}(2018)]%
        {mtmap2}
\bibfield{author}{\bibinfo{person}{Tatsuro Yamada}, \bibinfo{person}{Hiroyuki Matsunaga}, {and} \bibinfo{person}{Tetsuya Ogata}.} \bibinfo{year}{2018}\natexlab{}.
\newblock \showarticletitle{Paired Recurrent Autoencoders for Bidirectional Translation Between Robot Actions and Linguistic Descriptions}.
\newblock \bibinfo{journal}{\emph{IEEE Robotics and Automation Letters}} \bibinfo{volume}{3}, \bibinfo{number}{4} (\bibinfo{year}{2018}), \bibinfo{pages}{3441--3448}.
\newblock


\bibitem[Zeng et~al\mbox{.}(2022)]%
        {zeng2022glm}
\bibfield{author}{\bibinfo{person}{Aohan Zeng}, \bibinfo{person}{Xiao Liu}, \bibinfo{person}{Zhengxiao Du}, \bibinfo{person}{Zihan Wang}, \bibinfo{person}{Hanyu Lai}, \bibinfo{person}{Ming Ding}, \bibinfo{person}{Zhuoyi Yang}, \bibinfo{person}{Yifan Xu}, \bibinfo{person}{Wendi Zheng}, \bibinfo{person}{Xiao Xia}, {et~al\mbox{.}}} \bibinfo{year}{2022}\natexlab{}.
\newblock \showarticletitle{Glm-130b: An open bilingual pre-trained model}.
\newblock \bibinfo{journal}{\emph{arXiv preprint arXiv:2210.02414}} (\bibinfo{year}{2022}).
\newblock


\bibitem[Zhang et~al\mbox{.}(2023b)]%
        {T2M-GPT}
\bibfield{author}{\bibinfo{person}{Jianrong Zhang}, \bibinfo{person}{Yangsong Zhang}, \bibinfo{person}{Xiaodong Cun}, \bibinfo{person}{Shaoli Huang}, \bibinfo{person}{Yong Zhang}, \bibinfo{person}{Hongwei Zhao}, \bibinfo{person}{Hongtao Lu}, {and} \bibinfo{person}{Xi Shen}.} \bibinfo{year}{2023}\natexlab{b}.
\newblock \showarticletitle{T2M-GPT: Generating Human Motion from Textual Descriptions with Discrete Representations}. In \bibinfo{booktitle}{\emph{Proceedings of the IEEE/CVF Conference on Computer Vision and Pattern Recognition (CVPR)}}.
\newblock


\bibitem[Zhang et~al\mbox{.}(2019)]%
        {zhang2019bertscore}
\bibfield{author}{\bibinfo{person}{Tianyi Zhang}, \bibinfo{person}{Varsha Kishore}, \bibinfo{person}{Felix Wu}, \bibinfo{person}{Kilian~Q Weinberger}, {and} \bibinfo{person}{Yoav Artzi}.} \bibinfo{year}{2019}\natexlab{}.
\newblock \showarticletitle{Bertscore: Evaluating text generation with bert}.
\newblock \bibinfo{journal}{\emph{arXiv preprint arXiv:1904.09675}} (\bibinfo{year}{2019}).
\newblock


\bibitem[Zhang et~al\mbox{.}(2023a)]%
        {zhang2023motiongpt}
\bibfield{author}{\bibinfo{person}{Yaqi Zhang}, \bibinfo{person}{Di Huang}, \bibinfo{person}{Bin Liu}, \bibinfo{person}{Shixiang Tang}, \bibinfo{person}{Yan Lu}, \bibinfo{person}{Lu Chen}, \bibinfo{person}{Lei Bai}, \bibinfo{person}{Qi Chu}, \bibinfo{person}{Nenghai Yu}, {and} \bibinfo{person}{Wanli Ouyang}.} \bibinfo{year}{2023}\natexlab{a}.
\newblock \showarticletitle{MotionGPT: Finetuned LLMs are General-Purpose Motion Generators}.
\newblock \bibinfo{journal}{\emph{arXiv preprint arXiv:2306.10900}} (\bibinfo{year}{2023}).
\newblock


\bibitem[Zhao et~al\mbox{.}(2023)]%
        {Zhao_2023_CVPR}
\bibfield{author}{\bibinfo{person}{Qitao Zhao}, \bibinfo{person}{Ce Zheng}, \bibinfo{person}{Mengyuan Liu}, \bibinfo{person}{Pichao Wang}, {and} \bibinfo{person}{Chen Chen}.} \bibinfo{year}{2023}\natexlab{}.
\newblock \showarticletitle{PoseFormerV2: Exploring Frequency Domain for Efficient and Robust 3D Human Pose Estimation}. In \bibinfo{booktitle}{\emph{Proceedings of the IEEE/CVF Conference on Computer Vision and Pattern Recognition (CVPR)}}. \bibinfo{pages}{8877--8886}.
\newblock


\end{thebibliography}

\appendix

\section{Implementation Details}
\label{appendix}

In this Appendix section, we provide details about the data structure prepared for different language models about the action description task, the hyperparameters used, and the computations of different metrics.

\subsection{Preparing Data for Tuning Action Description Models}

    For language models that we evaluated for the action description task stated in Section~\ref{comparison}, namely the T5 family \cite{2020t5}, ChatGLM2 6B \cite{du2022glm,zeng2022glm}, and Llama 2 7B \cite{touvron2023llama}, the \textit{raw} textual data we built using outputs of VQ-VAE ($[\mathbf{Q}]$) and action recognition module ($[y]$), and physios' annotations are stated in Section~\ref{tuning} as:

    $\bullet$ Task prompt $\mathcal{T}$ : \{\textit{I want you to act as an action interpreter. Given the type of human action and tokens representing the action, please generate a natural language description of the action.}\}

    $\bullet$ Condition input: \{\textit{The action you need to describe is as following, type:} $[y]$, \textit{tokens:} $[\mathbf{Q}]$.\} 

    Such data was directly transformed to fit the training/tuning scheme of these language models. For ChatGLM2 6B, following the instruction provided in its official repository\footnote{ChatGLM2 6B with P-tuning, \url{https://github.com/THUDM/ChatGLM2-6B/tree/main/ptuning}} about customized fine-tuning with P-tuning v2 \cite{liu2022ptuning}, the data was transformed into the structure as shown in Algorithm~\ref{glm_input}. In particular, English was used during data preparation for all the models. Since English contents are also part of the pre-training for ChatGLM2, this practice shall not pose noticeable disadvantage on its fine-tuning given our data. For T5 small, base, large, 3B models, the same data structure as ChatGLM2 6B was used. For Llama 2 7B, we used the tuning framework presented in this repository\footnote{LLaMA-Factory: \url{https://github.com/hiyouga/LLaMA-Factory/tree/main}}, and our data was shaped as shown in Algorithm~\ref{llama_input}.

\begin{algorithm}[ht]
\SetAlgoLined
\texttt{\{\\
"instruction": "I want you to act as an action interpreter. Given the type of human action and tokens representing the action, please generate a natural language description of the action. The action you need to describe is as following, type: [y], tokens:[Q].",\\
"response": "[GROUND TRUTH]"\\
\}}
\caption{Data Structure for Fine-Tuning ChatGLM2 6B and T5 models}
\label{glm_input}
\end{algorithm}

\begin{algorithm}[ht]
\SetAlgoLined
\texttt{"columns": \{\\
    "prompt": "I want you to act as an action interpreter. Given the type of human action and tokens representing the action, please generate a natural language description of the action.",\\
    "query": "The action you need to describe is as following, type: [y], tokens:[Q].",\\
    "response": "[GROUND TRUTH]"\\
  \}}
\caption{Data Structure for Fine-Tuning Llama 2 7B}
\label{llama_input}
\end{algorithm}

\subsection{Hyperparameters}

    For VQ-VAE, the size of the codebook $\mathcal{B}$ is $512\times512$, the downsampling rate is set to $l=4$. Based on empirical evidence, $\beta$ is set to $1.0$ for the commitment loss, and $\alpha$ is set to 0.5 for the regularization loss using biomechanical features. Typical VQ-VAE training techniques such as codebook reset and exponential moving average (EMA) \cite{razavi2019generating} are used, where the exponential moving constant is set as $\lambda=0.99$. During training, the batch size is set to 128, the default AdamW optimizer \cite{loshchilov2017decoupled} is used for optimization, the initial learning rate is set to 2e-4 and is decreased by multiplying 0.1 after the first 100K steps, and the total number of steps is 150K. A non-overlapped sliding window is applied to the action feature inputs of different lengths. The window length is set to 64 for most experiments, although we also conducted an ablation experiment on this hyperparameter, which is reported in the experimental section.

    For all the language models, AdamW \cite{loshchilov2017decoupled} is used as the optimizer, and the pre-trained weights are all loaded before fine-tuning. For T5 models, the initial learning rate is set to 1e-4, and the batch size is 16. Particularly, T5 3B is fine-tuned using LoRA\cite{hu2021lora}, with $\alpha_{lora} = 16$, $dropout_{lora} = 0.1$, $rank_{lora}=64$. For Llama 2 7B, the initial learning rate is set to 3e-3 with a cosine learning rate scheduler, and the batch size is 64, 8-bit quantization with single precision is used, with the input length set to 512. LoRA is also used for fine-tuning Llama 2 7B, with $\alpha_{lora} = 16$, $dropout_{lora} = 0.1$, $rank_{lora}=64$. For ChatGLM2 6B, the initial learning rate is set to 5e-3, and the batch size is 16, the single precision is used without quantization. Here, P-tuning v2 \cite{liu2022ptuning} is used for its fine-tuning, with soft prompt length set to $128$, input length set to $512$. The architecture of the two convolutional backbones used in VQ-VAE and action recognition module are reported in the following tables, respectively.

\begin{table}[ht]
\label{table5}
\caption{The architecture of the convolutional backbone used in VQ-VAE. The exact input dimension of 315 was used for the input of biomechanical features, which was simply switched to 287 for action features adopted in previous works.}
\resizebox{0.71\linewidth}{!}{\begin{tabular}{cl}
Components & \multicolumn{1}{c}{Architecture} \\
\hline
{Encoder} & \begin{tabular}[c]{@{}l@{}}(0): Conv1d(315, 512,   kernel\_size=(3,), stride=(1,), padding=(1,))\\      (1): ReLU()\\      (2): 2 $\times$ Sequential(\\      \quad(0): Conv1d(512, 512, kernel\_size=(4,), stride=(2,),   padding=(1,))\\      \quad(1): Resnet1D(\\      \quad\quad(model): Sequential(\\      \quad\quad\quad(0): 3 $\times$ ResConv1DBlock(\\      \quad\quad\quad\quad(norm1): Identity()\\      \quad\quad\quad\quad(norm2): Identity()\\      \quad\quad\quad\quad(activation1): ReLU()\\      \quad\quad\quad\quad(activation2): ReLU()\\      \quad\quad\quad\quad(conv1): Conv1d(512, 512, kernel\_size=(3,),   stride=(1,), padding=(9,), dilation=(9,))\\      \quad\quad\quad\quad(conv2): Conv1d(512, 512, kernel\_size=(1,),   stride=(1,))))))\\      (4): Conv1d(512, 512, kernel\_size=(3,), stride=(1,), padding=(1,))\end{tabular} \\
\hline
Codebook & nn.Parameter((512, 512), requires\_grad=False) \\
\hline
Decoder & \begin{tabular}[c]{@{}l@{}}(0): Conv1d(512, 512,   kernel\_size=(3,), stride=(1,), padding=(1,))\\      (1): ReLU()\\      (2): 2 $\times$ Sequential(\\      \quad(0): Resnet1D(\\      \quad\quad\quad(0): 3 $\times$ ResConv1DBlock(\\      \quad\quad\quad\quad(norm1): Identity()\\      \quad\quad\quad\quad(norm2): Identity()\\      \quad\quad\quad\quad(activation1): ReLU()\\      \quad\quad\quad\quad(activation2): ReLU()\\      \quad\quad\quad\quad(conv1): Conv1d(512, 512, kernel\_size=(3,),   stride=(1,), padding=(9,), dilation=(9,))\\      \quad\quad\quad\quad(conv2): Conv1d(512, 512, kernel\_size=(1,),   stride=(1,))))\\      \quad(1): Upsample(scale\_factor=2.0, mode='nearest')\\      \quad(2): Conv1d(512, 512, kernel\_size=(3,), stride=(1,),   padding=(1,)))\\      (4): Conv1d(512, 512, kernel\_size=(3,), stride=(1,), padding=(1,))\\      (5): ReLU()\\      (6): Conv1d(512, 315, kernel\_size=(3,), stride=(1,), padding=(1,))\end{tabular}
\end{tabular}}
\end{table}

\begin{table}[ht]
\label{table6}
\caption{The of the convolutional backbone used in the action recognition module.}
\resizebox{\linewidth}{!}{\begin{tabular}{ll}
Components &
  Architecture \\
  \hline
{Action   Type Classifier} &
  \begin{tabular}[c]{@{}l@{}}(0): Conv1d(315, 64,   kernel\_size=(7,), stride=(2,), padding=(3,))\\      (1): BatchNorm1d(64, eps=1e-05, momentum=0.1, affine=True,   track\_running\_stats=True)\\      (2): ReLU()\\      (3): MaxPool1d(kernel\_size=3, stride=2, padding=1, dilation=1, ceil\_mode=False)\\      (4): 2 $\times$ Sequential(\\      \quad\quad(0): 2 $\times$ ResidualBlock(\\      \quad\quad\quad\quad(1): Conv1d(64, 64, kernel\_size=(3,), stride=(1,),   padding=(1,))\\      \quad\quad\quad\quad(2): BatchNorm1d(64, eps=1e-05, momentum=0.1, affine=True,   track\_running\_stats=True)\\      \quad\quad\quad\quad(3): ReLU()\\      \quad\quad\quad\quad(4): Conv1d(64, 64, kernel\_size=(3,), stride=(1,),   padding=(1,))\\      \quad\quad\quad\quad(5): BatchNorm1d(64, eps=1e-05, momentum=0.1,   affine=True, track\_running\_stats=True)\\      \quad\quad\quad\quad(6): Dropout(p=0.5, inplace=False)\\      \quad\quad\quad\quad(7): Identity()))\\      (5): AdaptiveAvgPool1d(output\_size=1)\\      (6): Linear(in\_features=128, out\_features=25, bias=True)\end{tabular}
\end{tabular}}
\end{table}

\subsection{Metrics Computation}

    As mentioned in Section~\ref{implementation}, benchmark metrics like Bleu \cite{papineni2002bleu}, Rouge \cite{lin2004rouge}, Cider \cite{vedantam2015cider}, and BertScore \cite{zhang2019bertscore} are adopted to measure the quality of action descriptions. Details of the calculation are reported as follows: 

    $\bullet$ Bleu score was calculated with the python library NLTK (\texttt{nltk.translate.bleu\_score.sentence\_bleu}), while the geometric sequence smoothing from National Institute of Standards and Technology (NIST) that is provided in this same package (\texttt{nltk.translate.bleu\_score.SmoothingFunction.method3}) was used as the smoothing function. 

    $\bullet$ ROUGE score was calculated with python library Rouge. No other additional parameters were set.

    $\bullet$ CIDEr score was calculated with python library Pycocoevalcap (\texttt{pycocoevalcap.cider}). No other additional parameters were set.

    $\bullet$ BertScore was calculated with official python library Bert-Score, with the following arguments: \texttt{lang='en', verbose=True, rescale\_with\_baseline=True, idf=True}. The default model for English, \texttt{roberta-large}, was employed to compute the scores we have presented.
    
\end{document}